\documentclass[sigconf]{acmart}
\usepackage{CJKutf8}
\usepackage{multirow}
\usepackage{multicol}

\AtBeginDocument{%
  \providecommand\BibTeX{{%
    \normalfont B\kern-0.5em{\scshape i\kern-0.25em b}\kern-0.8em\TeX}}}

\setcopyright{acmcopyright}
\copyrightyear{2018}
\acmYear{2018}
\acmDOI{XXXXXXX.XXXXXXX}

\acmConference[Conference acronym 'XX]{Make sure to enter the correct
  conference title from your rights confirmation emai}{June 03--05,
  2018}{Woodstock, NY}
\acmPrice{15.00}
\acmISBN{978-1-4503-XXXX-X/18/06}

\begin{document}
\begin{CJK}{UTF8}{gbsn}
\title{Multi-view Semantic Matching of Question retrieval using Fine-grained Semantic Representations}


\author{Li Chong}
\affiliation{%
  \institution{Renmin University of China}
  \city{Beijing}
  \country{China}}
\email{chongli@ruc.edu.cn}

\author{Denghao Ma}
\affiliation{%
  \institution{Meituan}
  \city{Beijing}
  \country{China}}
\email{madenghao5@gmail.com}

\author{Yueguo Chen}
\affiliation{%
  \institution{Renmin University of China}
  \city{Beijing}
  \country{China}}
\email{chenyueguo@ruc.edu.cn}

\renewcommand{\shortauthors}{Chong and Ma, et al.}

\begin{abstract}
As a key task of question answering, question retrieval has attracted much attention from the communities of academia and industry. 
Previous solutions mainly focus on the translation model, topic model, and deep learning techniques. Distinct from the previous solutions, we propose to construct fine-grained semantic representations of a question by a learned importance score assigned to each keyword, so that we can achieve a fine-grained question matching solution with these semantic representations of different lengths.
Accordingly, we propose a multi-view semantic matching model by reusing the important keywords in multiple semantic representations. 

As a key of constructing fine-grained semantic representations, we are the first to use a cross-task weakly supervised extraction model that applies question-question labelled signals to supervise the keyword extraction process (i.e. to learn the keyword importance). The extraction model integrates the deep semantic representation and lexical matching information with statistical features to estimate the importance of keywords. We conduct extensive experiments on three public datasets and the experimental results show that our proposed model significantly outperforms the state-of-the-art solutions.
\end{abstract}




\maketitle

\section{Introduction}
\label{Introduction}
\begin{figure*}[t!]
\centering
\includegraphics[width=1.96\columnwidth]{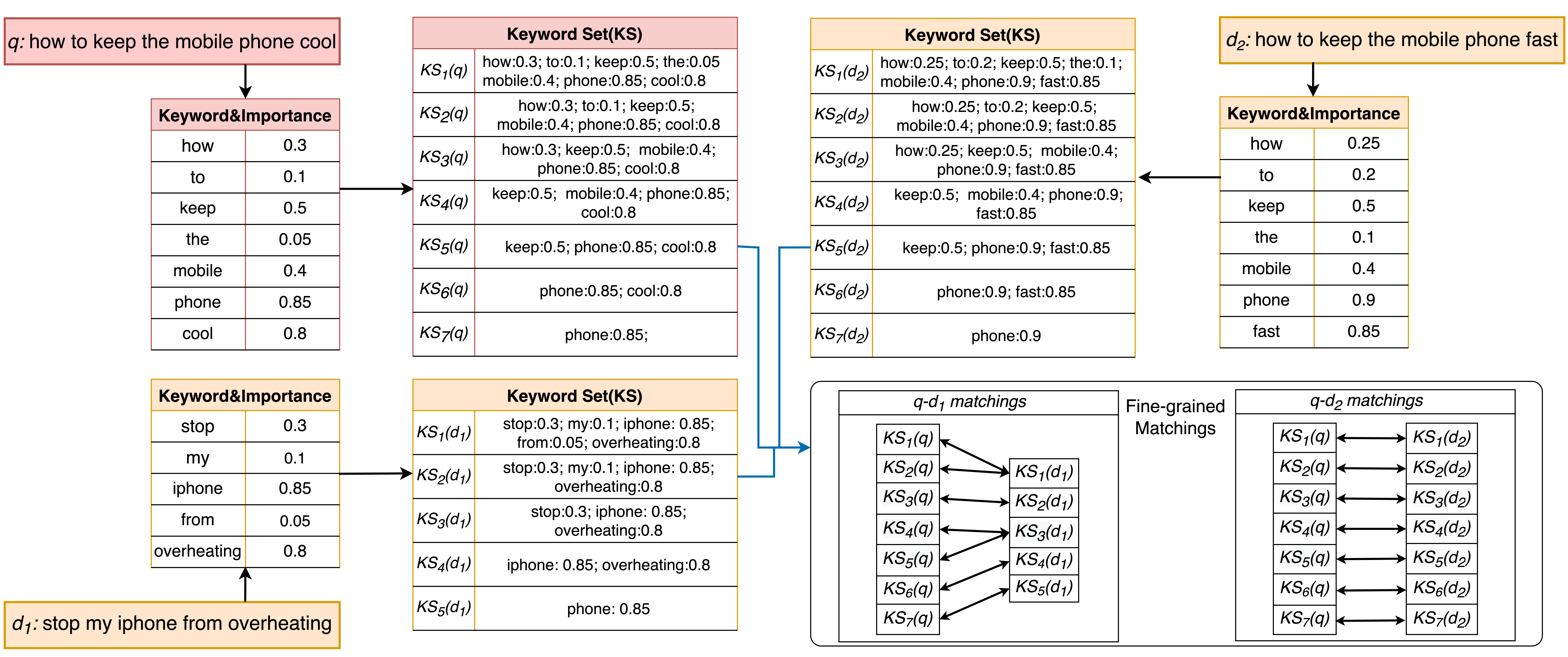}\vspace{-2mm}
\caption{Insights of reusing important keywords to construct the fine-grained representations and matchings.}
\label{Headword-based matching solutions} 
\vspace{-3mm}
\end{figure*}

Question retrieval is to retrieve semantically equivalent questions from an archived repository with large quantities of questions. It has been verified very important to both community-based question answering (CQA) \cite{DBLP:conf/semeval/NakovHMMMBV17}, e.g.,  \textit{Yahoo! Answers}, \textit{Quora} and \textit{Baidu Knows}, and domain-specific question answering, e.g., \textit{Apple Siri} and \textit{Microsoft Azure Bot}. 
Although many researchers from the communities of academia and industry have paid much attention to the task, it still suffers from a key challenge, i.e., the lexical gap.  The lexical gap contains two aspects: 1) textually distinct yet semantically equivalent; 2) textually similar yet semantically distinct. 
For example in Figure \ref{Headword-based matching solutions}, given three questions $q$=``how to keep the mobile phone
cool'', $d_1$=``stop my iphone from overheating'' and $d_2$=``how to keep the mobile phone fast'', $q$ and $d_1$ are textually distinct yet semantically equivalent, while $q$ and $d_2$ are textually similar yet semantically distinct.
The lexical gap hinders the effective retrieval of semantically equivalent questions from the archives to a user's query (a question or some keywords).

To address the challenge of textually distinct yet semantically equivalent, many solutions have been proposed based on the translation model \cite{DBLP:conf/naacl/MurdockC05}, topic model \cite{DBLP:conf/cikm/JiXWH12}, and deep learning techniques\cite{DBLP:conf/sigir/SakataSTK19}. 
The solutions based on translation model \cite{DBLP:conf/naacl/MurdockC05,DBLP:conf/acl/ZhouCZL11} learn translation probabilities for both words and phrases over some parallel corpora, and then use the translation probabilities to estimate the semantic similarity of two questions. 
The solutions based on topic model  \cite{DBLP:conf/ijcnlp/CaiZLZ11,DBLP:conf/cikm/JiXWH12,DBLP:conf/acl/PeineltNL20} construct the latent topic distributions for questions and estimate the semantic similarity of the two questions based on their topic distributions. 
Deep learning solutions mainly focus on 1) modelling the semantic relations of two questions by different neural network architectures and then estimating their similarity based on the relations \cite{DBLP:conf/sigir/SeverynM15,DBLP:conf/acl/YangZGJC19,DBLP:conf/taln/PontesHLT18}; 2) constructing good semantic representations for questions, such as BERT \cite{DBLP:conf/naacl/DevlinCLT19}  and ERNIE \cite{DBLP:conf/aaai/SunWLFTWW20}. These solutions model the semantics of a question by using a global representation, i.e., a bag of words, topic distribution and embedding, and globally match two questions. Just because of the ``global representation'', questions with similar text yet distinct semantics have similar topic distribution representation and embedding representation; Because of the ``globally match'', the above solutions hardly distinguish the subtle differences of important keywords of two questions and thus can not effectively address the challenge of textually similar yet semantically distinct.

To address the two aspects of the lexical gap, this paper attempts to find answers for the following two research questions: 1) How
to represent questions? 2) How to match two questions?

\textbf{Insight one: multi-level keyword sets.}
We propose new insights of reusing important keywords to construct fine-grained semantic representations of questions and then fine-grained matchings. 
For example in Figure \ref{Headword-based matching solutions}, given two questions $q$=``how to keep the mobile phone cool'' and $d_1$=``stop my iphone from overheating'', their corresponding keywords can be extracted and assigned different importance scores; Based on the importance scores, the multi-level keyword sets can be generated, i.e., the question $q$ has $7$ keyword sets and $d_1$ has $5$ keyword sets; For $q$, $KS_{1}(q)$ represents the finest semantic representation of $q$, and $KS_{7}(q)$ represents the coarsest one of $q$. 
The multi-level keyword sets can model the question semantics of various granularity, i.e., from ``finest''  to ``coarsest''. 
The finest keyword set can be used for modelling the global semantics of a question and then addressing the challenge of textually distinct yet semantically equivalent.
The other sets are reusing the important keywords so that the subtle differences of important keywords of two questions can be distinguished and then the challenge of textually similar yet semantically distinct can be addressed by the fine-grained matchings. For example in Figure \ref{Headword-based matching solutions}, for the question $q$ and $d_2$, the matchings between $KS_5(q)$ and $KS_5(d_2)$ as well as $KS_6(q)$ and $KS_6(d_2)$ can effectively identify the semantic distinction between $q$ and $d_2$.

\textbf{Insight two: comparable keyword set pairs.} Given two questions and their multi-level keyword sets, how to match them? For example in Figure \ref{Headword-based matching solutions},  the question $q$ and $d_1$ have $7$ and $5$ keyword sets respectively; We match $KS_1(q)$ to $KS_1(d_1)$, $KS_2(q)$ to $KS_1(d_1)$, $KS_3(q)$ to $KS_2(d_1)$, $KS_4(q)$ to $KS_3(d_1)$, $KS_5(q)$ to $KS_3(d_1)$, $KS_6(q)$ to $KS_4(d_1)$ and $KS_7(q)$ to $KS_5(d_1)$, because each pair is at the same or similar semantic level and is comparable. Reversely, $KS_1(q)$ and $KS_5(d_1)$ as well as $KS_7(q)$ and $KS_1(d_1)$ are not comparable because they are at different semantic level. The matchings on comparable keyword set pairs can benefit the similarity estimation of two questions, while that on the incomparable keyword set pairs will hurt the similarity estimation. Therefore, to effectively estimate the similarity of two questions, we need to construct the comparable keyword set pairs for the two questions.

\textbf{Solution: fine-grained matching network.} Based on the insights, we propose the fine-grained matching network with two cascaded units to estimate the similarity of two questions. \\
\textit{Fine-grained Representation Unit.}
According to the insight one, the fine-grained representation unit  constructs the multi-level keyword sets of a question for modelling its fine-grained semantics. 
Because of the deficiency of keyword annotations, we are the first to use the question-question labeled signals from the training set of question retrieval to supervise the keyword extraction (i.e., learn the keyword importance), and develop a cross-task weakly supervised  extraction model. In the model, we integrate deep semantic representations, lexical information, part-of-speech and statistical features to estimate the importance of keywords in a question. Based on the importance, the multi-level keyword sets are generated by iteratively removing one keyword with the lowest importance. \\
\textit{Fine-grained Matching Unit.}
According to the insight two, we design a pattern-based assignment method to construct  the comparable keyword set pairs from the multi-level keyword sets of two questions. The fine-grained matching unit estimates the semantic similarity of two questions by matching their comparable keyword set pairs.
First, it matches two questions from the global matching to local matching by using the comparable pairs of different granularity. Second, it matches every comparable pair from both the semantic matching and lexical matching perspectives. 
For the semantic matching, we develop two types of methods, i.e., MLP-based matching and attention-based matching; For the lexical matching, we use some widely used textual matching models such as BM25 and the Jaccard similarity. Third, it aggregates these matchings of multiple comparable pairs by learning their weights, and outputs an overall score as the semantic similarity of two questions.

Our contributions are concluded as follows:\\
\noindent$\bullet$ We propose new insights of reusing important keywords to construct the fine-grained representations and matchings, and design the fine-grained matching network to estimate the similarity of two questions from multiple granularities and multiple views.\\
\noindent$\bullet$ We are the first to use question-question labeled signals to supervise the keyword extraction process and develop the cross-task weakly supervised extraction model. \\
\noindent$\bullet$ We propose a pattern-based assignment method to construct the comparable keyword set pairs. \\
\noindent$\bullet$ We conduct extensive experiments on three public datasets and validate the effectiveness of the fine-grained matching network.

\section{Related Work}
Community-based question answering (CQA) portals have been popular platforms in which users may
ask their questions and share their answers. Because of the opening and sharing mechanism, CQA portals have accumulated a large quantity of questions and answers, which has made CQA portals valuable resources. 
As the quantity increases, users' questions may be repeated or closely related to previously archived questions which may be replied with some answers. Therefore, CQA portals first retrieve semantically equivalent questions
(i.e. question retrieval) from the archived repository and then identify the
high-quality answers (i.e. answer selection) of these
equivalent questions. So question retrieval is an important task in CQA portals. 

The question retrieval task suffers from the lexical gap challenge, and many solutions have been proposed to address it. We survey these solutions and group them into three types:\\
\noindent\textbf{Solutions based on translation model.} These solutions define the similarity of two questions as the translation probability from one to the other \cite{DBLP:conf/cikm/JeonCL05, DBLP:conf/sigir/XueJC08}. The study \cite{DBLP:conf/sigir/XueJC08} proposes a retrieval model consisting of a translation-based language model for the question
part and  a query likelihood approach for the answer part.
The proposed model incorporates word-to-word translation
probabilities learned from different parallel corpora. The study \cite{DBLP:conf/acl/ZhouCZL11} proposes a phrase-based translation model for question retrieval.

\noindent\textbf{Solutions based on topic model.} These solutions \cite{DBLP:conf/ijcnlp/CaiZLZ11,DBLP:conf/cikm/JiXWH12,DBLP:conf/cikm/ZhangWWLZ14} define the similarity in the latent topic space. The work \cite{DBLP:conf/ijcnlp/CaiZLZ11} proposes a topic model that incorporates category information into the process of discovering latent topics. The work \cite{DBLP:conf/cikm/JiXWH12} proposes a Question-Answer Topic Model (QATM) that learns the latent topics from the question-answer pairs, by assuming that a question and its paired answer share the same topic distribution. In the work \cite{DBLP:conf/cikm/ZhangWWLZ14}, a topic-based language model is proposed to match questions not only at a term level but also at a topic level.

\noindent\textbf{Deep learning solutions.}
The deep learning techniques have been widely applied in question retrieval, and achieve better performance than traditional solutions. 
The work \cite{DBLP:conf/sigir/SeverynM15}
designs a convolutional neural network that learns an optimal representation of question pairs and a similarity function to relate them in a supervised way from the available training data.
The work \cite{DBLP:conf/aaai/WanLGXPC16} uses a bidirectional LSTM to generate multiple positional sentence representations for each question. In the work \cite{DBLP:conf/acl/DasYCS16}, the Siamese
CNN is proposed to estimate the semantic similarity of two questions.

Recently, the pre-training language
models have demonstrated strong performance on
text representations in many NLP tasks, such as BERT \cite{DBLP:conf/naacl/DevlinCLT19}, Sentence-BERT \cite{DBLP:conf/emnlp/ReimersG19}  and ERNIE \cite{DBLP:conf/aaai/SunWLFTWW20}. Meanwhile, many models are proposed based on various pre-trained language models to estimate the similarity of two questions. 
The work \cite{DBLP:conf/sigir/SakataSTK19} presents a BERT-based FAQ retrieval system which uses BERT to estimate the correlation between a question and an answer. 
The work \cite{DBLP:conf/acl/MassCRK20} develops a fully unsupervised method which uses question answer pairs to train two BERT models and uses the BERT models to
match user queries to answers and questions, respectively. The work \cite{DBLP:conf/acl/PeineltNL20} proposes a topic-informed BERT-based architecture
to estimate the similarity of the two texts. In work \cite{DBLP:conf/bionlp/RawatWMRS20}, the medical entity information is learned via ERNIE and incorporated into some similarity estimation models.

\section{Fine-grained  Matching Network}

In this section, we first formulate the question retrieval problem,
and subsequently introduce our proposed fine-grained matching network with two cascaded units, i.e., fine-grained representation unit  and fine-grained matching unit. 

\subsection{Problem and The Overall Solution}
Question retrieval is to find semantically equivalent questions from the archived repository for a new question issued by a user. It mainly consists of two stages: 1) the recalling stage retrieves a subset of question candidates from the whole question repository. At this stage, many solutions have been proposed based on text matching and semantic matching;
2) The ranking stage sorts the candidates to generate the final results by using some fine-grained matching methods. In this paper, we focus on the ranking stage. Similar to many studies \cite{DBLP:conf/ijcai/WangHF17,DBLP:conf/coling/LiuCDZCLT18}, we assume that the question candidates exist. Given a new question $q$ and the candidates $D$, we formulate the semantic similarity of each $d \in D$ to $q$ as follows:
\begin{equation}
\label{p(q|d)}
sim(q, d)=f(q, d, \Phi)
\end{equation}
where the function $f$ takes $q$, $d$ and a set of parameters $\Phi$ as input and outputs a score ranging from $0$ to $1$ as the similarity between $q$ and $d$. According to  the scores, the candidates in $D$ can be ranked in descending order.
To design an effective function $f$, we need to answer two research problems--- how to represent  questions and how to match the two questions.

As mentioned in Section \ref{Introduction}, to answer the above two questions, we propose the fine-grained matching network with two cascaded units. The fine-grained representation unit uses the multi-level keyword sets to represent the fine-grained semantics of a question. The fine-grained matching unit first constructs multiple comparable  pairs of keyword sets for two questions, and then matches two questions from multiple granularities and multiple perspectives by using comparable keyword set pairs, i.e., from global matching to local matching and from lexical matching to semantic matching. The two units are simultaneously optimized by feedforward and backward propagation.  Given a question $q$ and a candidate question $d$, we formulate the network as follows:
\begin{equation}
\label{f_q_d_phi}
f(q, d, \Phi) = MLP(w_{s} E(S(q,d)) + b_{s})
\end{equation}
where $S(q,d)$ denotes the multiple comparable pairs of keyword sets and $E(S(q,d))$ is the multi-view matching representation over $S(q,d)$, which are constructed in Section \ref{Fine-grained Matching Unit}. The $MLP$ with parameters $w_s$ and $b_s$ is a $m\times n \times 1$ multi-layer perceptron where the activation functions of the first two layers are ReLU \cite{DBLP:journals/corr/ClevertUH15} and that of the third layer is Sigmoid \cite{DBLP:conf/softcomp/SharmaC11}. 

\subsection{Fine-grained Representation Unit}
\label{Fine-grained Representation Unit}
As motivated in Section \ref{Introduction}, to construct the multi-level keyword sets of a question, we need to extract and score keywords for a question. However, it is hard to apply a supervised model to extract and score keywords due to the lack of keyword annotations. Simply applying unsupervised models suffers from the precision of keyword extraction, which will be verified in the experiments. These motivate us to use question-question labeled
signals to supervise the keyword extraction process, leading to a cross-task weakly supervised extraction strategy.

\textit{Feasibility of the cross-task weakly supervised strategy.} On the one hand, accurate keyword extraction can benefit the similarity estimation.
For example in Figure \ref{Headword-based matching solutions}, given the user's question ``how to keep the mobile phone cool'' and the relevant archived question ``stop my iphone from overheating'', if the keyword extracting can effectively extract their corresponding keywords [``how'':$0.3$, ``to'':$0.1$,  ``keep'':$0.5$, ``the'':$0.05$, ``mobile'':$0.4$,  ``phone'': $0.85$,  ``cool'':$0.8$] and [``stop'':$0.3$, ``my'':$0.1$, ``iphone'':$0.85$, ``from'':$0.05$,  ``overheating'': $0.8$], the fine-grained matchings will be able to effectively estimate their similarity. 
Reversely, inaccurate keyword extraction will hurt the similarity estimation.
On the other hand, keywords with importance scores and fine-grained matching scores are propagated to the loss function, and the loss can be backward propagated to keyword extraction component. So the keyword extraction component can be optimized during the model training stage. 
To this end, in the fine-grained matching network, the importance of keywords is applied to the fine-grained matching unit. Consequently, we can use the question-question labeled signals to weakly supervise the keyword extracting and scoring.

\textit{Importance estimation.}
Given a question $q$ or $d$, we first segment the question to generate keyword candidates $\{k_1, \cdots, k_n\}$. It is trivial for an English-like question corpus. For the Chinese question corpus, we can use some commonly used segmentation tools, e.g., pkuseg \footnote{https://github.com/lancopku/pkuseg-python}, to segment each question, and take these segments as keyword candidates. For example, given a Chinese question  ``{\small 为啥我的苹果手机会发热}/why does my iphone overheat'', the segmentations are [``{\small 为啥}/why'',  ``{\small 我的}/my'', ``{\small 苹果手机}/iphone'', ``{\small 会}/does'', ``{\small 发热}/overheat'']. The importance of keywords in a question can be different to represent the meaning of the question. For the above example, the keyword candidate ``{\small 苹果手机}/iphone'' and ``{\small 发热}/overheat'' are more important to represent the semantics of the question than other ones. Therefore, we need to estimate the importance of a keyword $k_i$ to $q$, and develop the importance estimation network as follows:
\begin{equation}
\label{equation_confNN}
w(k_i, q)= ImNN(k_i, q)
\end{equation}
where $ImNN$ is a neural network that takes $k_i$ and  $q$ as input, and outputs an importance score ranging from $0$ to $1$.  We design the network $ImNN$ with three subnets: 1) $EmbNN$ learns the embedding representations for $k_i$ and $q$, and constructs their co-action embedding representations; 2) $EFNN$ constructs the external features of $k_i$; 3) $JINN$ processes the joint information from the previous two subnets and outputs the importance score. 

In $EmbNN$,  we use BERT \cite{DBLP:conf/naacl/DevlinCLT19} to generate the embedding of each candidate keyword. For Chinese questions, we first use BERT to generate the embedding $E_{w}$ for each character $w \in k_i$, and then use the embedding pooling of characters in $k_i$ as the embedding of  $k_i$, i.e., $E_{k_i}=\sum_{w \in k_i} E_{w}/|k_i|$ where $|k_i|$ is the textual length of $k_i$. For the question $q$, the embedding of $[CLS]$ is taken as the semantic embedding of $q$ (denoted as $E_{CLS}$). Based on $E_{k_i}$ and $E_{CLS}$, we formulate $EmbNN$ as follows:
\begin{equation}
\overline{embf}= MLP(w_e[E_{CLS} \oplus E_{k_i} \oplus CoA(E_{CLS},E_{k_i})] +b_e)
\end{equation}
\begin{equation}
\label{CoA}
CoA(E_{CLS},E_{k_i})= \sigma_{o}(w_o [E_{CLS}\oplus E_{k_i} \oplus|E_{CLS}-E_{k_i}| +b_o)
\end{equation}
where $\oplus$ is the concatenating operation and $|\cdot|$ is a sign of computing the absolute value. The $\sigma_o$ is the ReLU activation function and $MLP$ is a multi-layer perceptron with three layers whose activation functions are ReLU. The $w_*$ and $b_*$ are the weight matrix and  the bias vector.

The subnet $EFNN$ is to construct the external features for the keyword $k_i$. The inverse document frequency (IDF)--- the inverse fraction of the documents that contain the word, measures the quantity of information the word provides. To improve the generalization of our model, we split all IDF values into $K$ buckets and initialize each bucket into a learnable embedding with $64$ dimensions. In our experiments, $K=100$ by default. Therefore, the IDF value of $k_i$ is represented by an embedding, i.e., $EIDF_{k_i}$. The part of speech (POS) feature has been verified  important to many NLP tasks \cite{DBLP:conf/paclic/SuzukiKSS18,DBLP:conf/acl-sighan/LiZL17}. We use the pyltp tool \footnote{https://github.com/HIT-SCIR/pyltp} to mark the part of speech of each keyword candidate in Chinese sentences, and the Stanford POS tagger \footnote{https://nlp.stanford.edu/software/tagger.shtml} for English sentences. We train each POS into a learnable embedding, and denote the POS embedding of $k_i$ as  $EPOS_{k_i}$. Besides, we construct the stop-word feature for each word with $0$ or $1$ value, by using some stop-word dictionaries \footnote{https://github.com/goto456/stopwords; https://gist.github.com/sebleier/554280}. We denote the stop-word feature of $k_i$ as $ESW_{k_i}$.  Based on the $EIDF_{k_i}$, $EPOS_{k_i}$ and $ESW_{k_i}$ features, we formulate $EFNN$ as follows:
\begin{equation}
\overline{ef}= MLP(w_{f}[EIDF_{k_i} \oplus EPOS_{k_i} \oplus ESW_{k_i}]+b_f)
\end{equation}
where MLP is a multi-layer perceptron with three layers whose activation functions are ReLU. 

In the $JINN$ subnet, we first concatenate the embedding-based features $\overline{embf}$ and the external features $\overline{ef}$, and then use an MLP with three layers to estimate the importance of $k_i$ to $q$. Based on the $EmbNN$, $EFNN$ and $JINN$ subnets, we formulate the neural network $ImNN$ in Equation \ref{equation_confNN} as follows:
\begin{equation}
\label{confNN}
ImNN(k_i, q)= MLP(w_{j}[\overline{embf} \oplus \overline{ef}]+b_{j})
\end{equation}
where MLP is a multi-layer perceptron with three layers. The activation functions of the first two layers are ReLU, and that of the last layer is Sigmoid.

According to the keywords and their importance, we  remove one keyword with the lowest importance from the keyword set one by one, and each removal generates a keyword set. For example in Figure \ref{Headword-based matching solutions}, given question $q=$``how to keep the mobile phone cool'' and its keywords [``how'': $0.3$, ``to'':$0.1$; ``keep'': $0.5$, ``the'':$0.05$, ``mobile'':$0.4$, ``phone'': $0.85$, ``cool'':0.8], the highest level keyword set is $KS_1(q)$; Removing  ``the'': $0.05$ from $KS_1(q)$ generates the $KS_2(q)$, and  then removing ``to'': $0.1$ from $KS_2(q)$ generates the $KS_3(q)$; The $KS_1(q)$ is the finest semantic representation, and $KS_7(q)$ is the coarsest semantic representation. 

\subsection{Fine-grained Matching Unit}
\label{Fine-grained Matching Unit}
\subsubsection{Comparable Keyword Set Pairs}
As motivated in Section \ref{Introduction}, the matchings should be performed on the comparable keyword set pairs. So we formally define a comparable pair as follows:\\
\textit{\textbf{Comparable keyword set pair:} given two questions $q$ and $d$, and their multi-level keyword sets $KS(q)$ and $KS(d)$, a comparable keyword set pair is defined as $S_i=<KS_m(q), KS_n(d)>$ where $KS_m(q) \in KS(q)$ and $KS_n(d) \in KS(d)$ are two keyword sets that are at the same or similar semantic level.}

We develop the pattern-based assignment method to construct the comparable pair of keyword sets.  First, we compute the ratio that each keyword set covers the keywords in the highest level keyword set. 
For example in Figure \ref{Headword-based matching solutions}, given $q$ and the keyword sets $KS(q)=\{KS_1(q), \cdots, KS_7(q)\}$, the ratio of $KS_1(q)$ is $|KS_1(q)|/$ $|KS_1(q)|=1$, that of $KS_2(q)$ is $|KS_2(q)|$ / $|KS_1(q)|$ $=0.857$, and that of $KS_3(q)$ is $|KS_3(q)|/|KS_1(q)|=0.714$. 
Second, given two questions and their corresponding keyword sets and ratios, we extract keyword sets with the same or similar ratios as comparable pairs of keyword sets. Specifically, we use patterns to extract comparable keyword set pairs: given a user question $q$ and a keyword set $KS_m(q)$  with ratio $r_m(q)$, 1) if $KS_m(q)$ contains multiple keywords (i.e., $|KS_m(q)|>1$) and there is a keyword set $KS_n(d) \in KS(d)$ with the ratio $r_m(q)$, we record $KS_m(q)$ and $KS_n(d)$ as a comparable pair; 2) if $|KS_m(q)|>1$ and there is no keyword set in $KS(d)$ with the ratio $r_m(q)$, we extract the keyword set $KS_n(d)$ whose ratio is larger than $r_m(q)$ and is the closest to $r_m(q)$, and record $KS_n(d)$ and $KS_m(q)$ as a comparable pair; 3) if $|KS_m(q)|==1$, we use $KS_n(d)$ with only one keyword to pair $KS_m(q)$; 4) otherwise, we don't use $KS_m(q)$ to construct comparable pairs. By the pattern-based assignment method, a set of comparable pairs are constructed for the question $q$ and $d$. We denote the comparable pairs as $S(q,d)=\{\cdots,<KS_{m}(q), KS_{n}(d)>,\cdots\}$. 

\subsubsection{Multi-view Matching}
According to Equation \ref{f_q_d_phi}, we propose the multi-view matching over  $S(q,d)$ to construct $E(S(q,d))$:
\begin{equation}
E(S(q,d)) = \frac{\sum^{n}_{i=0} w(S_i(q,d)) mv(S_i(q,d))}{\sum^{n}_{i=0} w(S_i(q,d))}
\end{equation}
where $mv(S_i(q,d))$ is the multi-view matching on the pair $S_i(q,d)$. The $w(S_i(q,d))$ is the weight of  $S_i(q,d)$, because the matchings on the different comparable pairs have different importance. 
 
The pair $S_i(q,d)$ contains two keyword sets $KS_m(q)$ and $KS_n(d)$. If $KS_m(q)$ and $KS_n(d)$ are closer to the semantics of $q$ and $d$, the weight should be higher. 
In the keyword set, each keyword $k_i \in KS_{m}(q)$ (or $k_j \in KS_{n}(d)$) is assigned an importance. Based on the importance, we estimate the weight $w(S_i(q,d))$ as follows:
\begin{equation}
w(S_i(q,d)) = \frac{\sum_{k_i \in KS_{m}(q)}w(k_i,q)}{|q|} \cdot \frac{\sum_{k_j \in KS_{n}(d)}w(k_j,d)}{|d|} 
\end{equation}
where $w(k_i,q)$ is estimated by using Equation \ref{confNN}. The $|q|$ and $|d|$ are lengths of the question $q$ and $d$, respectively.

To effectively match the keyword set pair $KS_{m}(q)$ and $KS_{n}(d)$ ($S_i(q,d)$), we design a multi-view matching model from both semantic matching and lexical matching perspectives. The multi-view matching model is formulated as follows: 
\begin{equation}
\begin{aligned}
mv(S_i(q,d)) &=  m_{mlp}(KS_{m}(q), KS_{n}(d)) \\ 
 &\quad\oplus m_{att}(KS_{m}(q), KS_{n}(d)) \\ 
 &\quad\oplus m_{lm}(KS_{m}(q), KS_{n}(d))
\end{aligned}
\end{equation}
where $m_{mlp}$ is the MLP-based matching,  $m_{att}$ is the attention-based matching, and $m_{lm}$ matches them from the lexical perspective. We design the MLP-based matching $m_{mlp}$ as follows:
\begin{equation}
\begin{aligned}
\label{ms_mlp}
m_{mlp}(KS_{m}(q), KS_{n}(d)) &= \sigma(w_{m}[\overline{KS_{m}(q)}\oplus  \overline{KS_{n}(d)}\\ 
&\quad \oplus CoA(\overline{KS_{m}(q)}, \overline{KS_{n}(d)})]+ b_{m} )
\end{aligned}
\end{equation}
where $\overline{KS_{m}(q)}$ and $\overline{KS_{n}(d)}$ are the semantic representations of $KS_{m}(q)$ and $KS_{n}(d)$. The $CoA(\overline{KS_{m}(q)}, \overline{KS_{n}(d)})$ can be estimated by using Equation \ref{CoA}. 
We estimate $\overline{KS_{m}(q)}$ as follows:
\begin{equation}
\overline{KS_{m}(q)} = \sum_{k_i \in KS_{m}(q)} \frac{w(k_i, q)E_{k_i}}{\sum_{k_i \in KS_{m}(q)}w(k_i, q)}
\end{equation}
where $E_{k_i}$ is the embedding representation of $k_i$, which can be learned by some pre-trained language models. The $\overline{KS_{n}(d)}$ is constructed in the same way.

The MLP-based matching ignores the dependence of keywords.
So we design the attention-based matching to incorporate the dependence into the matching representation. 
Specifically, we concatenate the words in $KS_{m}(q)$ and $KS_{n}(d)$ separately, and denote the concatenated texts as $\hat{q}$ and $\hat{d}$. Secondly, we concatenate $\hat{q}$ and $\hat{d}$ by using $[CLS]$ to denote the start position and $[SEP]$ to differentiate them. We initialize the concatenated text with a list of embedding vectors $L$ and use the multi-head
scaled dot-production attention \cite{DBLP:conf/nips/VaswaniSPUJGKP17} to model the dependence among keywords:
\begin{equation}
head^{i}_L =softmax(\frac{LW^{i}_Q(LW^{i}_K)^{T}}{\sqrt{d_k}})LW^{i}_V
\end{equation}
\begin{equation}
head_L =[head^{1}_L, \cdots, head^{n}_L]W_o
\end{equation}
where $W^{i}_Q$, $W^{i}_K$, $W^{i}_V$ and $W_o$ are learnable weights. We use the position-wise feed forward
network (FFN) and layer normalization (LNorm) to further refine $head_L$: 
\begin{equation}
\label{ms_att}
m_{att}(KS_{m}(q), KS_{n}(d)) = LNorm(head_L+FFN(head_L))
\end{equation}
where $m_{att}$ is the attention-based matching representation.

Different from $m_{mlp}$ and $m_{att}$, $m_{lm}$ matches $KS_{m}(q)$ and $KS_{n}(d)$ from the lexical perspective. It uses multiple solutions to match $KS_{m}(q)$ and $KS_{n}(d)$. The function $m_{lm}$ is formulated as follows: 
\begin{equation}
\label{ms_lm}
m_{lm}(KS_{m}(q), KS_{n}(d))\!\! = BM25 \oplus  Jaccard \oplus wlm(KS_{m}(q), KS_{n}(d))
\end{equation}
\begin{equation}
wlm(KS_{m}(q), KS_{n}(d))\!\! =\!\!\!\!\sum_{k_i \in KS_{m}(q)}\! \sum_{k_j \in  KS_{n}(d)}\!\!\!\! w(k_i,q) w(k_j,d) a(k_i, k_j)
\end{equation}
where $BM25$ is the Okapi BM25 matching function \footnote{https://en.wikipedia.org//wiki//Okapi\_BM25
}, and $Jaccard$ is the Jaccard similarity coefficient \footnote{https://en.wikipedia.or//wiki//Jaccard\_index}. The $a(k_i, k_j)$ equals $1$ if $k_i = k_j$, otherwise $a(k_i, k_j)$ equals $0$. The $w(k_i,q)$ is the importance score of $k_i$ to $q$, and it is estimated by using Equation \ref{confNN}.

\subsubsection{Training and Inference}
At the training stage, we use the Log-likelihood loss to optimize the fine-grained matching network and present the loss function as follows:
\begin{equation}
 \mathcal{L} = -\frac{1}{|\mathcal{C}|} \sum_{<q, d, y> \in \mathcal{C} } \{y log(sim(q,d))+ (1-y) log(1-sim(q,d))\}
\end{equation}
where $\mathcal{C}$ is a set of training examples, and $|\mathcal{C}|$ is the size of $\mathcal{C}$. The $y$ is the ground truth label of $q$ and $d$, and $y=1$ denotes that $d$ is semantically equivalent  to $q$. The $sim(q,d)$ is the similarity between $q$ and $d$, which is estimated by using Equation \ref{p(q|d)}. At the inference stage, only if $sim(q,d) > 0.5$, $d$ is identified semantically equivalent to $q$. 
\section{Experiments}
\subsection{Experimental Setup}
\subsubsection{Datasets Description} We conduct experiments on three public datasets, show
the statistical information of the datasets in Table \ref{Statistical information of datasets}, and introduce them as follows:\\
$\bullet$ BankQ \cite{DBLP:conf/emnlp/ChenCLYLT18}: It is the largest dataset of question retrieval in the financial domain and sampled from the session logs of an online bank custom service system.
\\
\noindent$\bullet$ Quora\footnote{https://data.quora.com/First-Quora-Dataset-ReleaseQuestion-Pairs}: The dataset is sampled from the American Knowledge Q\&A website Quora.com.  Each question pair is labeled with a binary value indicating whether the two questions are similar.\\
\noindent$\bullet$ LCQMC \cite{DBLP:conf/coling/LiuCDZCLT18}: It is a large-scale Chinese question matching corpus and sampled from the largest online Chinese question answering platform, i.e.,  Baidu Knows.
\begin{table}[t]
	\centering
	\caption{Statistical information of datasets.\vspace{-2mm}}
		\label{Statistical information of datasets}
		\begin{tabular}{p{1.0cm}|p{1.2cm}p{1.2cm}p{1.0cm}p{1.0cm}}  
			\toprule
			Dataset & Type & Train & Dev & Test\\
			\midrule
			\multirow{2}{*}{BankQ}
			 &positive& 50000 & 5000 & 5000\\
			 &negative& 50000 & 5000 & 5000\\
			\hline
			\multirow{2}{*}{Quora}
			 &positive& 139306 & 5000 & 5000\\
			 &negative& 245042 & 5000 & 5000\\
			\hline
			\multirow{2}{*}{LCQMC}
			 &positive& 138574 & 4402 & 6250\\
			 &negative& 100192 & 4400 & 6250\\
           	\bottomrule
		\end{tabular}
		\vspace{-2mm}
\end{table}
\vspace{-1mm}
\subsubsection{Comparison Solutions.}
We select some state-of-the-art solutions  as baselines to verify the effectiveness of our model:\\
\noindent$\bullet$ BiMPM \cite{DBLP:conf/ijcai/WangHF17}: It first encodes the question $q$ and $d$ with a BiLSTM encoder, and then matches the two questions by using multiple matching methods, i.e., full-matching, max-pooling-matching, attentive-matching, and max-attentive-matching. 
\\
\noindent$\bullet$ RE2 \cite{DBLP:conf/acl/YangZGJC19}: It is a fast and strong neural architecture for general-purpose text matching. It  integrates the previously aligned features (Residual vectors), original point-wise features, and
contextual features as input of the inter-sequence alignment.\\
\noindent$\bullet$ BERT \cite{DBLP:conf/naacl/DevlinCLT19}: It is designed to pre-training deep bidirectional representations and achieves new state-of-the-art results on many natural language processing tasks including question retrieval.
\\
\noindent$\bullet$ ERNIE \cite{DBLP:conf/aaai/SunWLFTWW20}: It is a continual pre-training framework to effectively capture the lexical, syntactic, and semantic information from training corpora. It uses the constant multi-task learning strategy to  learn incrementally some pre-training tasks.
\\
\noindent$\bullet$ Sentence-BERT \cite{DBLP:conf/emnlp/ReimersG19}: Using the framework of the twin network model, it inputs different sentences into the parameter shared BERT model, and outputs the embedding vector for each sentence.\\
\noindent$\bullet$ tBERT \cite{DBLP:conf/acl/PeineltNL20}: It uses topic models, such as LDA and GSDMM, to improve BERT for pairwise semantic similarity detection.\\
We denote our proposed model as FRM.
\vspace{-1mm}
\subsubsection{Performance metrics.} Similar to the studies \cite{DBLP:conf/acl/PeineltNL20,DBLP:conf/aaai/SunWLFTWW20,DBLP:conf/acl/YangZGJC19}, we use accuracy and AUC. metrics to measure the effectiveness of all models, and introduce them as follows:\\
\noindent$\bullet$ Accuracy  (Acc.): It is a widely used metric for evaluating classification models, and measures the fraction of correct predictions. Based on the positives and negatives, Acc. can be  calculated as $Acc. = \frac{(TP+TN)}{TP+FP+TN+FN}$where $TP$= true positives, $TN$ = true negatives, $FP$ = false positives, and $FN$ = false negatives.

\noindent$\bullet$ Area Under Curve (AUC.)\cite{DBLP:journals/prl/Fawcett06}: AUC. measures the two-dimensional area under the ROC curve. It represents the probability that a positive example is positioned in front of a random negative example. 
\vspace{-5mm}
\subsubsection{Reproducibility.}
The parameters of all models are assigned the default values for fair comparisons. Specifically, the batch sizes of models on the BankQ and LCQMC datasets are $32$ and that of models on Quora datasets is $64$. All models are developed in Python $3.8$ and Pytorch $1.10$ development environment. We set the learning rate to $2e-5$ and use the warm-up learning rate method. We apply dropout before every fully-connected layer of all models and set the dropout probability as $0.3$. All models are deployed on the same computing services so that they have the same running conditions. In models, the parameters are optimized by the AdamW optimizer \cite{DBLP:journals/corr/abs-1711-05101}. For fair comparisons, we perform every solution five times, and choose the best model to evaluate on the test set.
\vspace{-1mm}
\subsection{Experimental Results}
\begin{table}[t]
	\centering
	\caption{Performance comparison over BankQ.\vspace{-2mm}}
		\label{Overall metric comparison over BankQ}
		
		\begin{tabular}{p{2.6cm}|p{1.0cm}p{1.0cm}} 
			\toprule
			Model & Acc.(\%) & AUC.(\%)\\
			\midrule
			BiMPM & 79.48  & 87.50 \\
			RE2   &  81.07 & 88.94  \\
			BERT   &  84.06  & 89.34\\
			ERNIE   &  84.66 & 92.31  \\
			Sentence-BERT   &  83.53 & 90.95 \\
			tBERT   &  80.65  &  88.82\\
            \hline
            FRM &  \textbf{85.16} & \textbf{92.43}\\
           	\bottomrule
		\end{tabular}
		\vspace{-2mm}
\end{table}

\begin{table}[t]
	\centering
	\caption{Performance comparison over LCQMC.\vspace{-2mm}}
		\label{Overall metric comparison over LCQMC}
		\begin{tabular}{p{2.6cm}|p{1.0cm}p{1.0cm}} 
			\toprule
			Model & Acc.(\%) & AUC.(\%)\\
			\midrule
			BiMPM & 83.59 & 93.75 \\
			RE2   &  84.74 & 94.78  \\
			BERT   & 86.70 & 94.84\\
			ERNIE   &  87.17 & 95.89 \\
			Sentence-BERT   & 84.07 & 94.62\\
			tBERT   &  85.12  &  93.13\\
            \hline
            FRM &  \textbf{89.05} & \textbf{96.17}\\
           	\bottomrule
		\end{tabular}
		\vspace{-2mm}
\end{table}

\begin{table}[t]
	\centering
	\caption{Performance comparison over Quora.\vspace{-2mm}}
		\label{Overall metric comparison over Quora}
		
		\begin{tabular}{p{2.6cm}|p{1.0cm}p{1.0cm}}  
			\toprule
			Model & Acc.(\%) & AUC.(\%)\\
			\midrule
			BiMPM & 85.82  & 93.46 \\
			RE2  &  89.20  & 95.56 \\
			BERT  &  91.07  & 96.96 \\
			ERNIE  &  91.08  & 96.84 \\
			Sentence-BERT  & 88.69 & 94.84\\
	        tBERT  &  89.31  & 95.69\\
            \hline
            FRM & \textbf{91.64}  & \textbf{97.01} \\
           	\bottomrule
		\end{tabular}
		\vspace{-2mm}
\end{table}

\begin{figure*}[t]
	\centering
	\begin{minipage}{0.35\linewidth}
		\centering
		\includegraphics[width=0.9\linewidth]{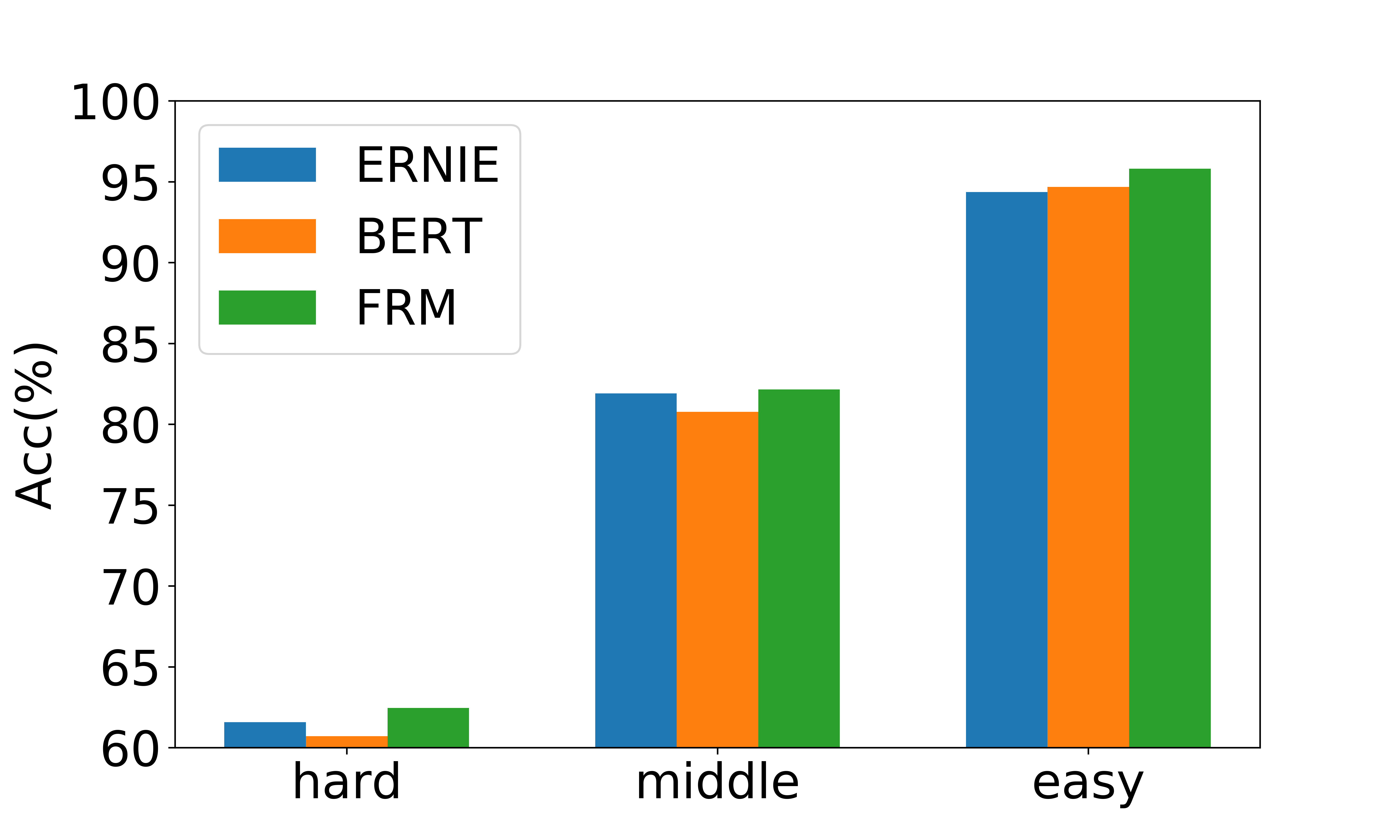}\vspace{-4mm}
		\caption*{(a) BankQ\vspace{-2mm}}
	\end{minipage}\hspace{-8mm}
	\begin{minipage}{0.35\linewidth}
		\centering
		\includegraphics[width=0.9\linewidth]{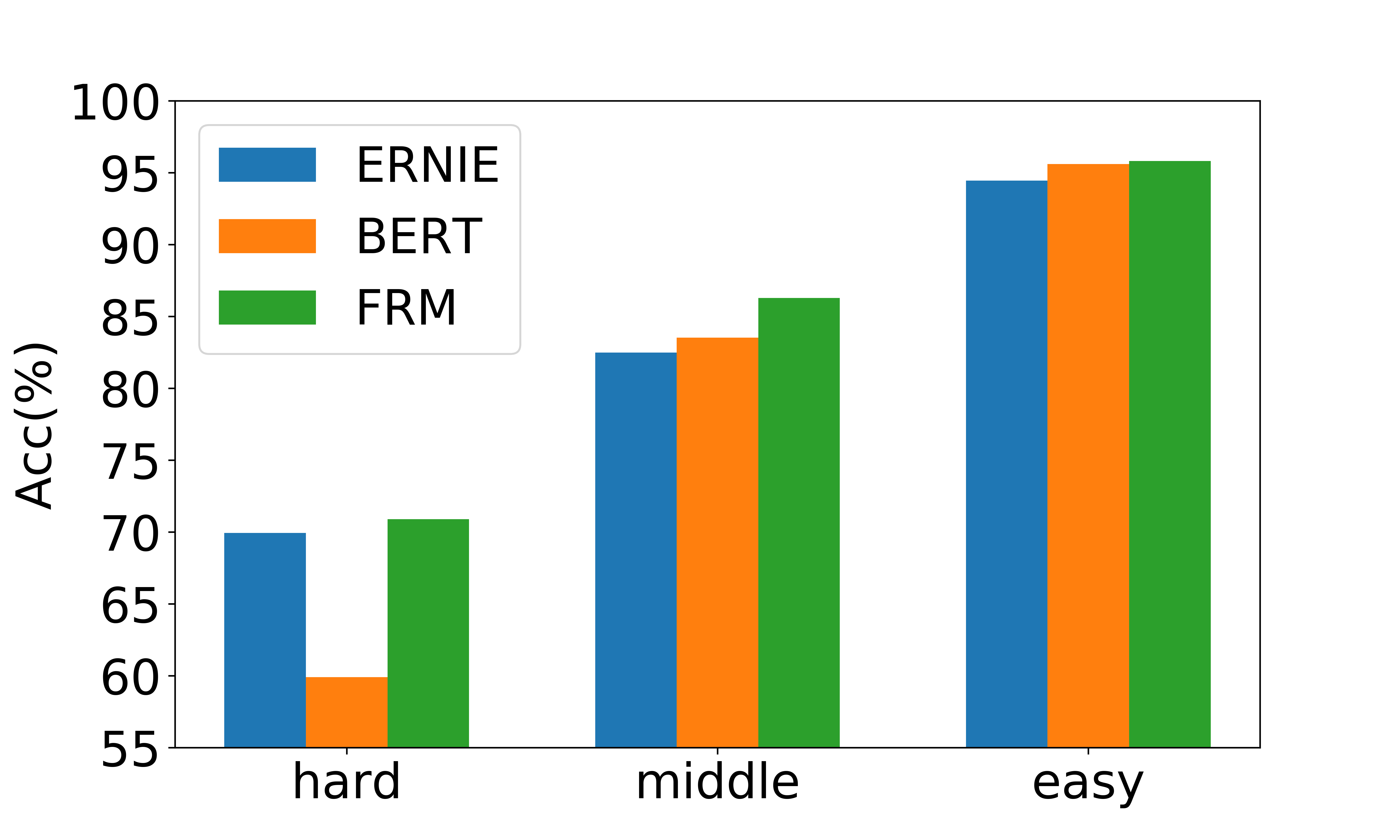}\vspace{-4mm}
		\caption*{(b) LCQMC\vspace{-2mm}}
	\end{minipage}\hspace{-8mm}
	\begin{minipage}{0.35\linewidth}
		\centering
		\includegraphics[width=0.9\linewidth]{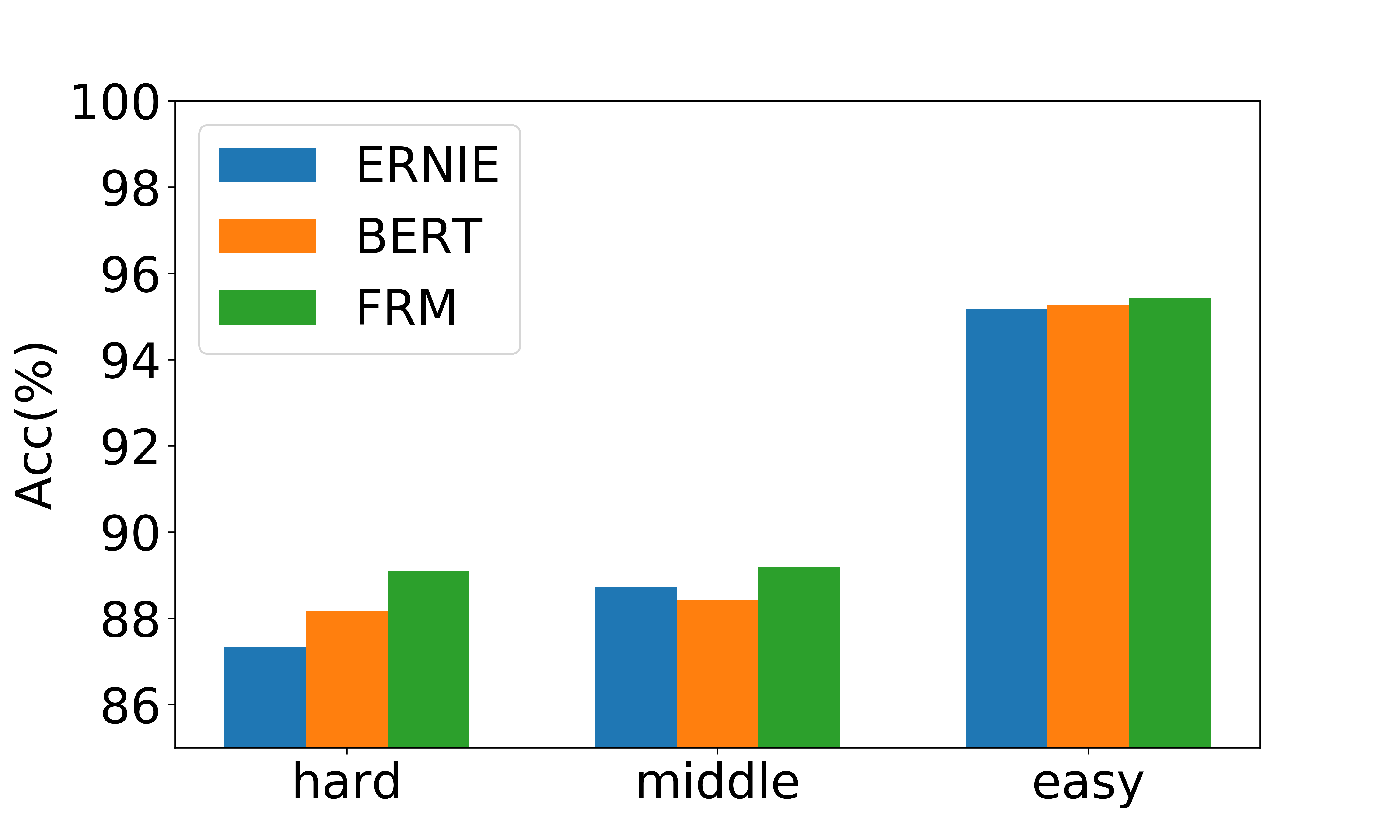}\vspace{-4mm}
		\caption*{(c) Quora\vspace{-2mm}}
		
	\end{minipage}
	\vspace{-1mm}
	\caption{The performance of FRM, BERT and ERNIE on the hard, middle and easy samples.\vspace{-2mm}}
	\label{The performance of all models on the hard, easy and other samples.}
	\vspace{-2mm}
\end{figure*}

\begin{figure*}[t]
	\centering
	\begin{minipage}{0.35\linewidth}
		\centering
		\includegraphics[width=0.9\linewidth]{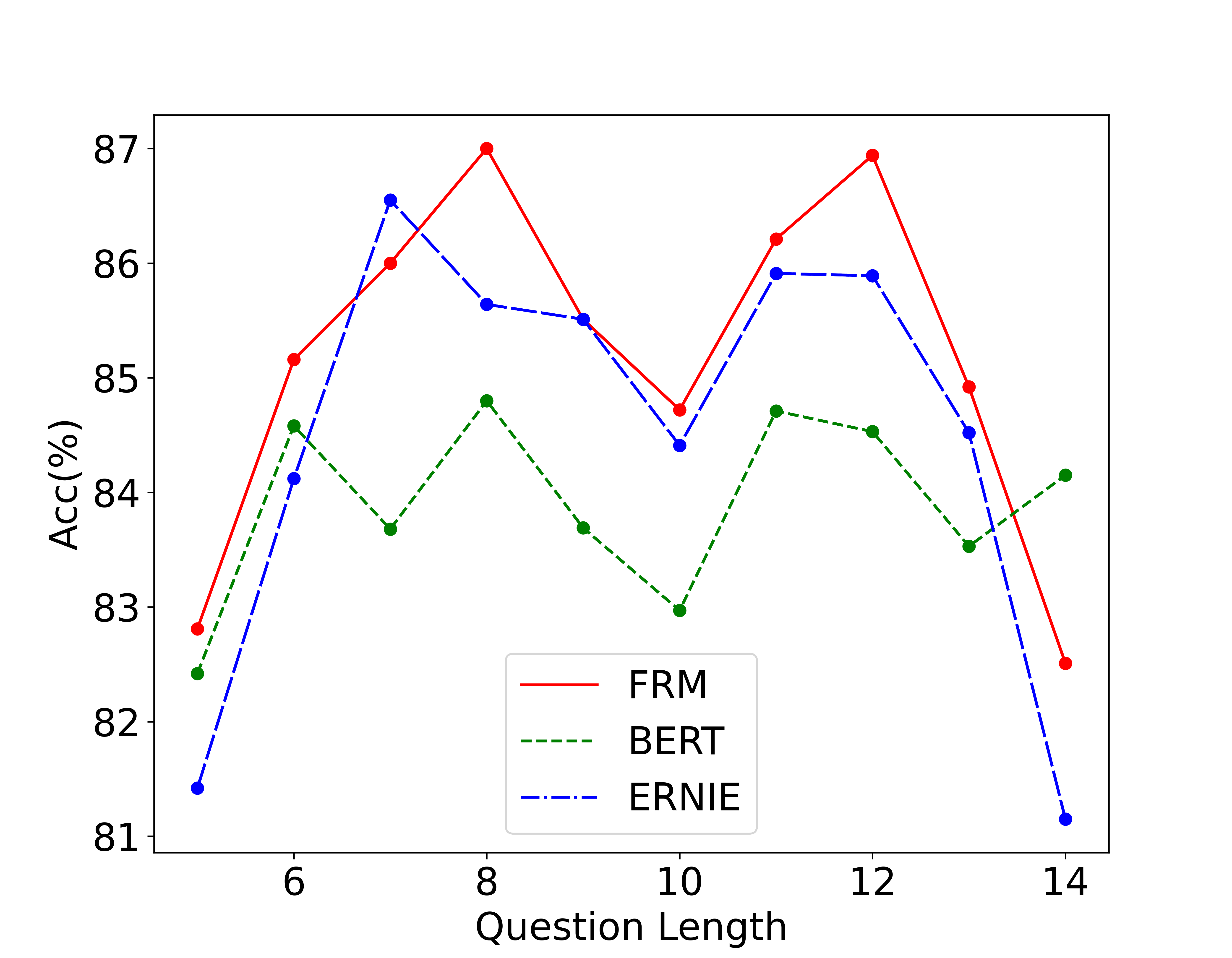}\vspace{-4mm}
		\caption*{(a) BankQ\vspace{-2mm}}
		
	\end{minipage}\hspace{-8mm}
	\begin{minipage}{0.35\linewidth}
		\centering
		\includegraphics[width=0.9\linewidth]{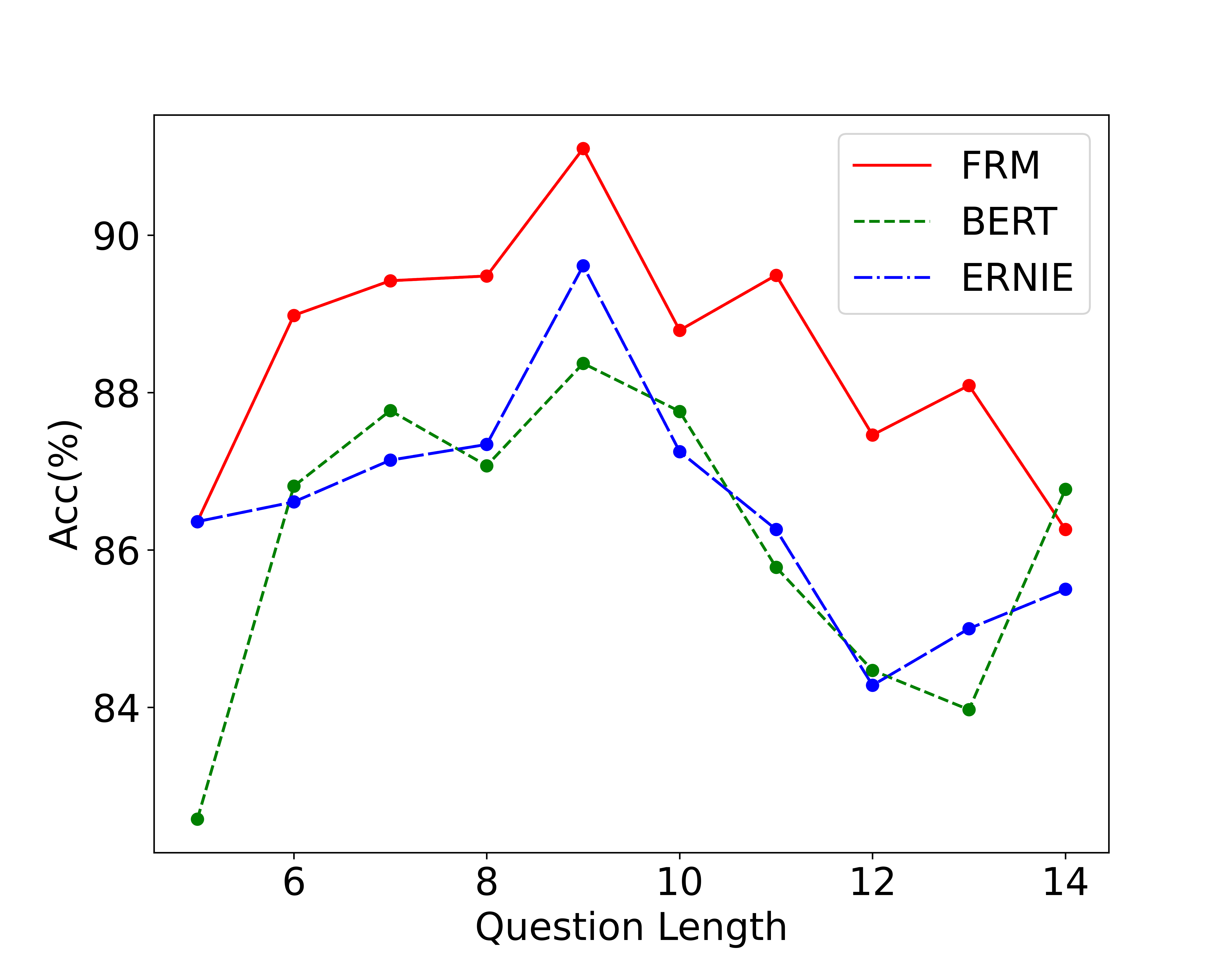}\vspace{-4mm}
		\caption*{(b) LCQMC\vspace{-2mm}}
		
	\end{minipage}\hspace{-8mm}
	\begin{minipage}{0.35\linewidth}
		\centering
		\includegraphics[width=0.9\linewidth]{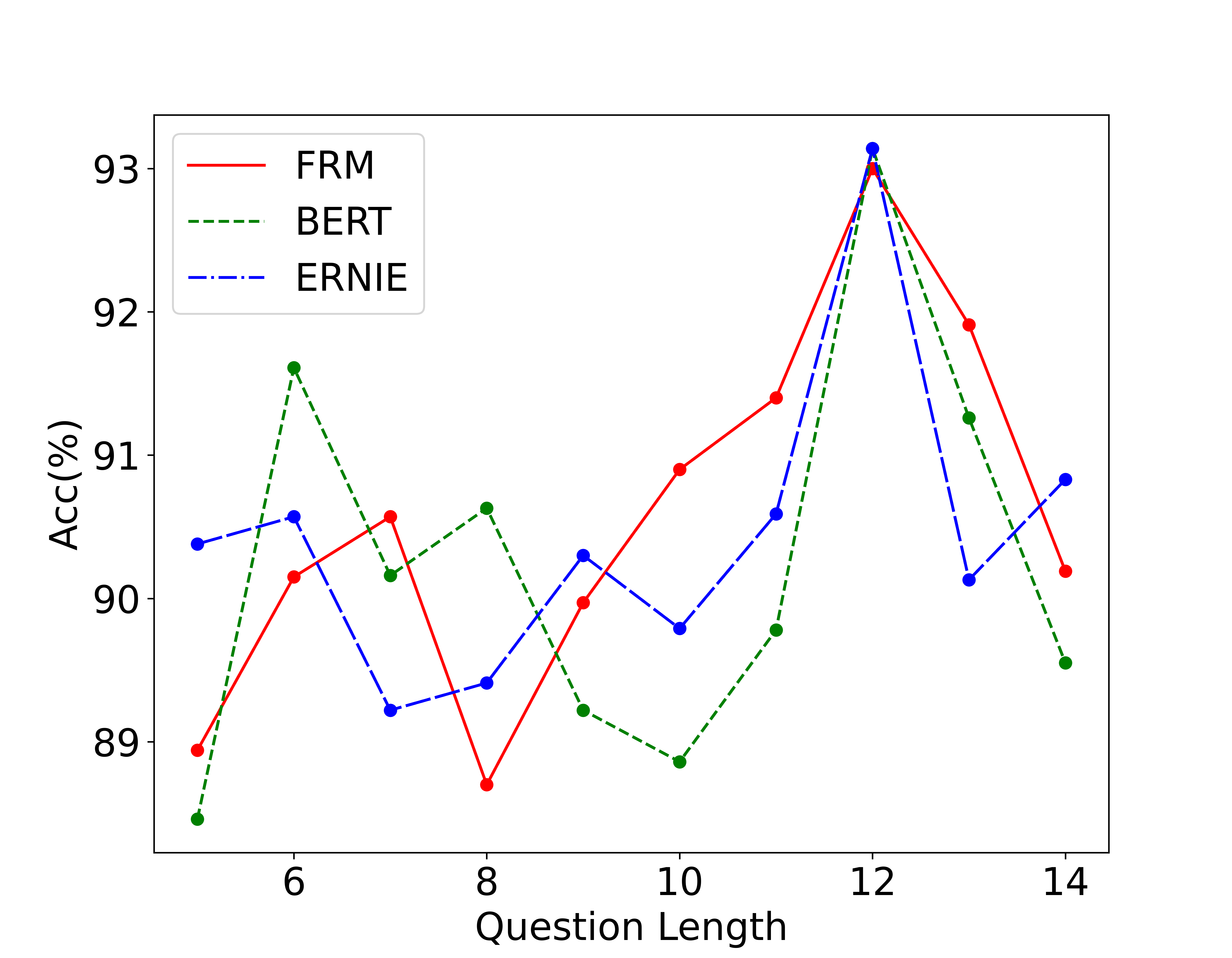}\vspace{-4mm}
		\caption*{(c) Quora\vspace{-2mm}}
		
	\end{minipage}
	\vspace{-1mm}
	\caption{The effect of question lengths on the performance of FRM, BERT and ERNIE.\vspace{-2mm}}
	\label{The effect of question length on the performance of FRM, BERT and ERNIE}
	\vspace{-2mm}
\end{figure*}

A summary of the performance comparison over three datasets is
displayed in Table \ref{Overall metric comparison over BankQ}, \ref{Overall metric comparison over LCQMC} and \ref{Overall metric comparison over Quora}. According to the results, it can be seen that 1) the performance achieved by the baselines BERT and ERNIE is higher than that achieved by the other baseline models. This verifies that BERT and ERNIE are very strong baseline models. Compare to BIMPM and RE2, BERT and ERNIE are pre-training representation models that benefit from the high-quality pre-training representations; 2) The performance achieved by FRM is better than that achieved by the best baseline models, i.e., BERT and ERNIE. Compared to BERT and ERNIE, FRM benefits from fine-grained representations and fine-grained matchings. The performance improvements verify the effectiveness of FRM.

To further verify the effectiveness of our model and the better baselines in addressing the lexical gap challenge (i.e., textually distinct yet semantically equivalent and textually similar yet semantically distinct), we first classify the test samples into the hard, easy and middle types; Second, we investigate the Acc. metrics of  models on the three types of samples, and report the results in Figure \ref{The performance of all models on the hard, easy and other samples.}. About the classification of samples, we group the positive samples of textually distinct yet semantically equivalent (i.e., the Jaccard similarity is in $[0.0,0.2]$) and the negative samples of textually similar yet semantically distinct (i.e., the Jaccard similarity is in $[0.7,1.0]$) into the hard type; The positive samples of textually similar and semantically equivalent (i.e., the Jaccard similarity is in $[0.7,1.0]$) and the negative samples of textually distinct and semantically distinct (i.e., the Jaccard similarity is in $[0.0,0.2]$) are grouped into the easy type; The remained samples are grouped into the middle type. 
In Figure \ref{The performance of all models on the hard, easy and other samples.}, it can be seen that 1) the Acc. values of all models on easy samples are higher than those on the middle samples, and the Acc. values of all models on middle samples are higher than those on the hard samples; 2) The FRM model performs better than BERT and ERNIE on the three types of samples. The advantage of FRM on the hard samples is more significant than that on easy and middle samples. These illustrate that FRM can better address the lexical gap challenge than BERT and ERNIE. 

We investigate the effect of question lengths on the performance of FRM, BERT and ERNIE and show the results in Figure \ref{The effect of question length on the performance of FRM, BERT and ERNIE}. It can be seen that 1) the Acc. values are lower when the question length is smaller or larger. This is because questions with few words can not provide rich semantics, and the models hardly  capture the semantic focus of questions with too many words; 2) Overall, FRM performs better than BERT and ERNIE. This illustrates that the advantage of FRM is not affected by the question lengths, even though the construction of multi-level keyword sets depends on the question words.

The fine-grained matching unit matches two questions from the perspectives of both semantic matching and lexical matching. To achieve semantic matching, we develop the MLP-based matching function $m_{mlp}$ (see Equation \ref{ms_mlp}) and  the attention-based matching function $m_{att}$ (see Equation \ref{ms_att}); To achieve lexical matching, we develop the matching function $m_{lm}$ (see Equation \ref{ms_lm}). To verify the effectiveness of the three matching functions and investigate their effects on metrics, we make an ablation study and present the results in Table \ref{The effectiveness of fine-grained matching unit}. The notation FRM-$m_{mlp}$, FRM-$m_{att}$ and FRM-$m_{lm}$ represent that the matching function $m_{mlp}$, $m_{att}$ and $m_{lm}$are not applied to FRM, respectively. According to the results in Table \ref{The effectiveness of fine-grained matching unit}, we can find that 1) FRM-$m_{mlp}$ achieves the lowest accuracy, which illustrates that $m_{mlp}$ is more important to the accuracy metric than the other two functions; 2) FRM-$m_{att}$ achieves the lowest AUC, which illustrates that $m_{att}$ is more important to the AUC metric than the $m_{mlp}$ and $m_{lm}$ functions; 3) Generally, the semantic matching functions ($m_{mlp}$ and $m_{att}$) are more important to the question retrieval task than the lexical matching function ($m_{lm}$). This is because lexical matching suffers from the severe lexical gap problem, while semantic matching can address the lexical gap to some extent. This also demonstrates the existence of the lexical gap problem. In addition, the performance achieved by FRM is much higher than that achieved by FRM-$m_{mlp}$, FRM-$m_{att}$ and FRM-$m_{lm}$. The performance improvement illustrates that the combination of semantic matching and lexical matching is better than any individual, and verifies the effectiveness of the multi-view matching. 

We improve the baseline BERT by incorporating the lexical matching (denoted as $BM25$) used in the fine-grained matching unit and the external features (denoted as $ef$) used in the fine-grained representation unit.  Specifically, given two questions, we first use BERT to encode them and use the embedding of $[CLS]$ to represent the encoded semantics ($E_{CLS}$). Second, we use BM25 model to estimate the similarity of the two questions and denote it as $BM25$. Third, we concatenate $E_{CLS}$ and $BM25$ as the final matching representation which is the input of a classification function. For the external features, we use them to extend the semantics of a word. The results are presented in Table \ref{The effectiveness of fine-grained matching unit}. The BERT+$ef$ and BERT+$BM25$ perform better than BERT, which provides an insight that we can use external features and lexical matching to improve BERT. Besides, the performance of FRM is still better than those of ERT+$ef$ and BERT+$BM25$. This illustrates that FRM exploits the external features and lexical matching more effectively than BERT+$ef$ and BERT+$BM25$. 

\begin{table}[t]
	\centering
	\caption{The effectiveness of fine-grained matching unit.\vspace{-2mm}}
		\label{The effectiveness of fine-grained matching unit}
		\begin{tabular}{p{2.8cm}|p{1.0cm}p{1.0cm}}  
			\toprule
			Model & Acc.(\%) & AUC.(\%)\\
			\midrule
			FRM-$m_{mlp}$  & 83.64 & 92.03 \\
			FRM-$m_{att}$   & 83.93  & 91.58 \\
			FRM-$m_{lm}$    & 84.71 &  91.79 \\
			\hline
			BERT+$ef$  & 84.27 & 92.05\\
            BERT+$BM25$ & 84.53 & 91.98\\
            BERT   &  84.06  & 89.34\\
            \hline
            FRM &  \textbf{85.16} & \textbf{92.43}\\
           	\bottomrule
		\end{tabular}
		\vspace{-3mm}
  
\end{table}

We propose the comparable pairs of keyword sets to support the fine-grained matching unit.
To verify the effectiveness of multiple comparable pairs, we perform a group of experiments where different numbers of comparable pairs are applied to FRM. The results are shown in Table \ref{The effectiveness of fine-grained representation unit}. The notation FRM-$one\;level$ represents that only the highest-level keyword set pair is  applied; FRM-$two\;level$ ( or FRM-$three\;level$) denotes that the keyword set pairs of the first two (or three) levels are applied. From  Table \ref{The effectiveness of fine-grained matching unit}, we can find that 1) FRM-$three\;level$ performs better than FRM-$two\;level$ and FRM-$one\;level$; 2) FRM-$two\;level$ performs better than FRM-$one\;level$; 3) FRM achieves the best metrics than FRM-$three\;level$, FRM-$two\;level$ and FRM-$one\;level$. In other words, the more keyword set pairs are applied, the higher the performance is achieved by FRM. These performance improvements illustrate that the comparable keyword set pairs can effectively support the multi-view matching component. 

\begin{table}[t]
	\centering
	\caption{{\small The effectiveness of fine-grained representation unit.\vspace{-2mm}}}
		\label{The effectiveness of fine-grained representation unit}
		\begin{tabular}{p{2.8cm}|p{1.0cm}p{1.0cm}}  
			\toprule
			Model & Acc.(\%) & AUC.(\%)\\
			\midrule
			FRM-$ef$ & 84.38 & 92.20\\
			FRM-$dsr$ & 84.26 & 92.29\\
			FRM-$1$ & 84.12 & 92.19\\
			FRM-$TF\-IDF$ & 83.91 & 92.18\\
            \hline
			FRM-$one\;level$ & 84.48 & 91.12\\
			FRM-$two\;level$ & 84.73 & 92.13 \\
			FRM-$three\;level$ & 84.86 & 92.12\\
			\hline
            FRM &  \textbf{85.16} & \textbf{92.43}\\
           	\bottomrule
		\end{tabular}
		\vspace{-5mm}
\end{table}

To verify the effectiveness of the fine-grained representation unit in FRM, we perform a group of experiments to investigate the effect of the deep semantic representation and external features on the performance, and present the results in Table \ref{The effectiveness of fine-grained representation unit}. The FRM-$ef$ denotes that the external features are not applied, and FRM-$dsr$ denotes that the deep semantic representation is not applied. Firstly, the performance achieved by FRM-$ef$ and FRM-$dsr$ is lower than that achieved by FRM. This illustrates that the combination of the deep semantic representation and the external features is better than any individual for estimating the importance of keywords. Secondly, FRM-$ef$ performs better than FRM-$dsr$, which indicates that the deep semantic representation is more important than external features. In addition, we use different methods to replace our proposed keyword extraction component and label the experimental results as FRM-${1}$ and FRM-$TFIDF$ in Table \ref{The effectiveness of fine-grained representation unit}. The FRM-${1}$ denotes that all keywords are assigned as equal importance, i.e. $1$, and the FRM-$TFIDF$ denotes that the $TFIDF$ value of a keyword is used for estimating the importance. We can see that 1) the performance achieved by FRM-$TFIDF$ is not higher than that achieved by FRM-$1$. This verifies that the unsupervised method can not accurately estimate the importance scores of keywords; 2) FRM performs much better than FRM-$TFIDF$ and FRM-${1}$, which further verifies the effectiveness of the fine-grained representation unit. 

According to the results in Table \ref{Overall metric comparison over BankQ}, \ref{The effectiveness of fine-grained matching unit} and \ref{The effectiveness of fine-grained representation unit}, we find a phenomenon that some ablation models of FRM perform not better than the best baseline ERNIE, such as FRM-$m_{mlp}$, FRM-$m_{att}$, FRM-$ef$ and FRM-$dsr$. The FRM-$m_{mlp}$ and FRM-$m_{att}$ remove the very important components from the fine-grained matching unit in $FRM$, i.e., MLP-based matching and attention-based matching. 
The FRM-$ef$ and FRM-$dsr$ remove the external features and the deep semantic representation feature from the fine-grained representation unit so that the multi-level keyword sets of high quality can not be constructed. So the phenomenon is in line with our expectations, and further demonstrates the importance of these components to FRM. 

\section{Conclusion}
To accurately estimate the semantic similarity of two questions, we present the two aspects of the lexical gap challenge (i.e., textually distinct yet semantically equivalent and textually similar yet semantically distinct), and propose new insights of reusing important keywords to construct the fine-grained semantic representations and fine-grained matchings.
To realize the insights, we use the multi-level keyword sets to model the question semantics of different granularities, propose the comparable pairs of keyword sets to support the multi-view matchings, and develop the fine-grained matching network.
By experiments, we find that 1) the multi-level keyword sets can capture and model both the global and local semantics of a question; 2) the combination of semantic matching and lexical matching is more effective than any single one.



\bibliographystyle{ACM-Reference-Format}
\bibliography{reference}

\begin{thebibliography}{30}


\ifx \showCODEN    \undefined \def \showCODEN     #1{\unskip}     \fi
\ifx \showDOI      \undefined \def \showDOI       #1{#1}\fi
\ifx \showISBNx    \undefined \def \showISBNx     #1{\unskip}     \fi
\ifx \showISBNxiii \undefined \def \showISBNxiii  #1{\unskip}     \fi
\ifx \showISSN     \undefined \def \showISSN      #1{\unskip}     \fi
\ifx \showLCCN     \undefined \def \showLCCN      #1{\unskip}     \fi
\ifx \shownote     \undefined \def \shownote      #1{#1}          \fi
\ifx \showarticletitle \undefined \def \showarticletitle #1{#1}   \fi
\ifx \showURL      \undefined \def \showURL       {\relax}        \fi
\providecommand\bibfield[2]{#2}
\providecommand\bibinfo[2]{#2}
\providecommand\natexlab[1]{#1}
\providecommand\showeprint[2][]{arXiv:#2}

\bibitem[Cai et~al\mbox{.}(2011)]%
        {DBLP:conf/ijcnlp/CaiZLZ11}
\bibfield{author}{\bibinfo{person}{Li Cai}, \bibinfo{person}{Guangyou Zhou},
  \bibinfo{person}{Kang Liu}, {and} \bibinfo{person}{Jun Zhao}.}
  \bibinfo{year}{2011}\natexlab{}.
\newblock \showarticletitle{Learning the Latent Topics for Question Retrieval
  in Community {QA}}. In \bibinfo{booktitle}{\emph{Fifth International Joint
  Conference on Natural Language Processing, {IJCNLP} 2011, Chiang Mai,
  Thailand, November 8-13, 2011}}. \bibinfo{publisher}{The Association for
  Computer Linguistics}, \bibinfo{pages}{273--281}.
\newblock
\urldef\tempurl%
\url{https://aclanthology.org/I11-1031/}
\showURL{%
\tempurl}


\bibitem[Chen et~al\mbox{.}(2018)]%
        {DBLP:conf/emnlp/ChenCLYLT18}
\bibfield{author}{\bibinfo{person}{Jing Chen}, \bibinfo{person}{Qingcai Chen},
  \bibinfo{person}{Xin Liu}, \bibinfo{person}{Haijun Yang},
  \bibinfo{person}{Daohe Lu}, {and} \bibinfo{person}{Buzhou Tang}.}
  \bibinfo{year}{2018}\natexlab{}.
\newblock \showarticletitle{The {BQ} Corpus: {A} Large-scale Domain-specific
  Chinese Corpus For Sentence Semantic Equivalence Identification}. In
  \bibinfo{booktitle}{\emph{{EMNLP}}}. \bibinfo{publisher}{Association for
  Computational Linguistics}, \bibinfo{pages}{4946--4951}.
\newblock


\bibitem[Clevert and et~al.(2016)]%
        {DBLP:journals/corr/ClevertUH15}
\bibfield{author}{\bibinfo{person}{Djork{-}Arn{\'{e}} Clevert} {and}
  \bibinfo{person}{et al.}} \bibinfo{year}{2016}\natexlab{}.
\newblock \showarticletitle{Fast and Accurate Deep Network Learning by
  Exponential Linear Units(ELUs)}. In \bibinfo{booktitle}{\emph{ICLR}}.
\newblock


\bibitem[Das et~al\mbox{.}(2016)]%
        {DBLP:conf/acl/DasYCS16}
\bibfield{author}{\bibinfo{person}{Arpita Das}, \bibinfo{person}{Harish
  Yenala}, \bibinfo{person}{Manoj~Kumar Chinnakotla}, {and}
  \bibinfo{person}{Manish Shrivastava}.} \bibinfo{year}{2016}\natexlab{}.
\newblock \showarticletitle{Together we stand: Siamese Networks for Similar
  Question Retrieval}. In \bibinfo{booktitle}{\emph{{ACL} {(1)}}}.
  \bibinfo{publisher}{The Association for Computer Linguistics}.
\newblock


\bibitem[Devlin et~al\mbox{.}(2019)]%
        {DBLP:conf/naacl/DevlinCLT19}
\bibfield{author}{\bibinfo{person}{Jacob Devlin}, \bibinfo{person}{Ming{-}Wei
  Chang}, \bibinfo{person}{Kenton Lee}, {and} \bibinfo{person}{Kristina
  Toutanova}.} \bibinfo{year}{2019}\natexlab{}.
\newblock \showarticletitle{{BERT:} Pre-training of Deep Bidirectional
  Transformers for Language Understanding}. In
  \bibinfo{booktitle}{\emph{{NAACL-HLT} {(1)}}}.
  \bibinfo{publisher}{Association for Computational Linguistics},
  \bibinfo{pages}{4171--4186}.
\newblock


\bibitem[Fawcett(2006)]%
        {DBLP:journals/prl/Fawcett06}
\bibfield{author}{\bibinfo{person}{Tom Fawcett}.}
  \bibinfo{year}{2006}\natexlab{}.
\newblock \showarticletitle{An introduction to {ROC} analysis}.
\newblock \bibinfo{journal}{\emph{Pattern Recognit. Lett.}}
  \bibinfo{volume}{27}, \bibinfo{number}{8} (\bibinfo{year}{2006}),
  \bibinfo{pages}{861--874}.
\newblock
\urldef\tempurl%
\url{https://doi.org/10.1016/j.patrec.2005.10.010}
\showDOI{\tempurl}


\bibitem[Jeon et~al\mbox{.}(2005)]%
        {DBLP:conf/cikm/JeonCL05}
\bibfield{author}{\bibinfo{person}{Jiwoon Jeon}, \bibinfo{person}{W.~Bruce
  Croft}, {and} \bibinfo{person}{Joon~Ho Lee}.}
  \bibinfo{year}{2005}\natexlab{}.
\newblock \showarticletitle{Finding similar questions in large question and
  answer archives}. In \bibinfo{booktitle}{\emph{Proceedings of the 2005 {ACM}
  {CIKM} International Conference on Information and Knowledge Management,
  Bremen, Germany, October 31 - November 5, 2005}},
  \bibfield{editor}{\bibinfo{person}{Otthein Herzog},
  \bibinfo{person}{Hans{-}J{\"{o}}rg Schek}, \bibinfo{person}{Norbert Fuhr},
  \bibinfo{person}{Abdur Chowdhury}, {and} \bibinfo{person}{Wilfried Teiken}}
  (Eds.). \bibinfo{publisher}{{ACM}}, \bibinfo{pages}{84--90}.
\newblock
\urldef\tempurl%
\url{https://doi.org/10.1145/1099554.1099572}
\showDOI{\tempurl}


\bibitem[Ji et~al\mbox{.}(2012)]%
        {DBLP:conf/cikm/JiXWH12}
\bibfield{author}{\bibinfo{person}{Zongcheng Ji}, \bibinfo{person}{Fei Xu},
  \bibinfo{person}{Bin Wang}, {and} \bibinfo{person}{Ben He}.}
  \bibinfo{year}{2012}\natexlab{}.
\newblock \showarticletitle{Question-answer topic model for question retrieval
  in community question answering}. In \bibinfo{booktitle}{\emph{21st {ACM}
  International Conference on Information and Knowledge Management, CIKM'12,
  Maui, HI, USA, October 29 - November 02, 2012}},
  \bibfield{editor}{\bibinfo{person}{Xue{-}wen Chen}, \bibinfo{person}{Guy
  Lebanon}, \bibinfo{person}{Haixun Wang}, {and} \bibinfo{person}{Mohammed~J.
  Zaki}} (Eds.). \bibinfo{publisher}{{ACM}}, \bibinfo{pages}{2471--2474}.
\newblock
\urldef\tempurl%
\url{https://doi.org/10.1145/2396761.2398669}
\showDOI{\tempurl}


\bibitem[Li et~al\mbox{.}(2017)]%
        {DBLP:conf/acl-sighan/LiZL17}
\bibfield{author}{\bibinfo{person}{Shuihua Li}, \bibinfo{person}{Xiaoming
  Zhang}, {and} \bibinfo{person}{Zhoujun Li}.} \bibinfo{year}{2017}\natexlab{}.
\newblock \showarticletitle{Chinese Answer Extraction Based on {POS} Tree and
  Genetic Algorithm}. In \bibinfo{booktitle}{\emph{Proceedings of the 9th
  {SIGHAN} Workshop on Chinese Language Processing, SIGHAN@IJCNLP 2017, Taipei,
  Taiwan, December 1, 2017}}, \bibfield{editor}{\bibinfo{person}{Yue Zhang}
  {and} \bibinfo{person}{Zhifang Sui}} (Eds.). \bibinfo{publisher}{Association
  for Computational Linguistics}, \bibinfo{pages}{30--36}.
\newblock
\urldef\tempurl%
\url{https://aclanthology.org/W17-6004/}
\showURL{%
\tempurl}


\bibitem[Liu et~al\mbox{.}(2018)]%
        {DBLP:conf/coling/LiuCDZCLT18}
\bibfield{author}{\bibinfo{person}{Xin Liu}, \bibinfo{person}{Qingcai Chen},
  \bibinfo{person}{Chong Deng}, \bibinfo{person}{Huajun Zeng},
  \bibinfo{person}{Jing Chen}, \bibinfo{person}{Dongfang Li}, {and}
  \bibinfo{person}{Buzhou Tang}.} \bibinfo{year}{2018}\natexlab{}.
\newblock \showarticletitle{{LCQMC:} {A} Large-scale Chinese Question Matching
  Corpus}. In \bibinfo{booktitle}{\emph{{COLING}}}.
  \bibinfo{publisher}{Association for Computational Linguistics},
  \bibinfo{pages}{1952--1962}.
\newblock


\bibitem[Loshchilov and Hutter(2017)]%
        {DBLP:journals/corr/abs-1711-05101}
\bibfield{author}{\bibinfo{person}{Ilya Loshchilov} {and}
  \bibinfo{person}{Frank Hutter}.} \bibinfo{year}{2017}\natexlab{}.
\newblock \showarticletitle{Fixing Weight Decay Regularization in Adam}.
\newblock \bibinfo{journal}{\emph{CoRR}}  \bibinfo{volume}{abs/1711.05101}
  (\bibinfo{year}{2017}).
\newblock
\showeprint[arXiv]{1711.05101}
\urldef\tempurl%
\url{http://arxiv.org/abs/1711.05101}
\showURL{%
\tempurl}


\bibitem[Mass et~al\mbox{.}(2020)]%
        {DBLP:conf/acl/MassCRK20}
\bibfield{author}{\bibinfo{person}{Yosi Mass}, \bibinfo{person}{Boaz Carmeli},
  \bibinfo{person}{Haggai Roitman}, {and} \bibinfo{person}{David Konopnicki}.}
  \bibinfo{year}{2020}\natexlab{}.
\newblock \showarticletitle{Unsupervised {FAQ} Retrieval with Question
  Generation and {BERT}}. In \bibinfo{booktitle}{\emph{Proceedings of the 58th
  Annual Meeting of the Association for Computational Linguistics, {ACL} 2020,
  Online, July 5-10, 2020}}, \bibfield{editor}{\bibinfo{person}{Dan Jurafsky},
  \bibinfo{person}{Joyce Chai}, \bibinfo{person}{Natalie Schluter}, {and}
  \bibinfo{person}{Joel~R. Tetreault}} (Eds.). \bibinfo{publisher}{Association
  for Computational Linguistics}, \bibinfo{pages}{807--812}.
\newblock
\urldef\tempurl%
\url{https://doi.org/10.18653/v1/2020.acl-main.74}
\showDOI{\tempurl}


\bibitem[Murdock and Croft(2005)]%
        {DBLP:conf/naacl/MurdockC05}
\bibfield{author}{\bibinfo{person}{Vanessa Murdock} {and}
  \bibinfo{person}{W.~Bruce Croft}.} \bibinfo{year}{2005}\natexlab{}.
\newblock \showarticletitle{A Translation Model for Sentence Retrieval}. In
  \bibinfo{booktitle}{\emph{{HLT/EMNLP} 2005, Human Language Technology
  Conference and Conference on Empirical Methods in Natural Language
  Processing, Proceedings of the Conference, 6-8 October 2005, Vancouver,
  British Columbia, Canada}}. \bibinfo{publisher}{The Association for
  Computational Linguistics}, \bibinfo{pages}{684--691}.
\newblock
\urldef\tempurl%
\url{https://aclanthology.org/H05-1086/}
\showURL{%
\tempurl}


\bibitem[Nakov et~al\mbox{.}(2017)]%
        {DBLP:conf/semeval/NakovHMMMBV17}
\bibfield{author}{\bibinfo{person}{Preslav Nakov}, \bibinfo{person}{Doris
  Hoogeveen}, \bibinfo{person}{Llu{\'{\i}}s M{\`{a}}rquez},
  \bibinfo{person}{Alessandro Moschitti}, \bibinfo{person}{Hamdy Mubarak},
  \bibinfo{person}{Timothy Baldwin}, {and} \bibinfo{person}{Karin Verspoor}.}
  \bibinfo{year}{2017}\natexlab{}.
\newblock \showarticletitle{SemEval-2017 Task 3: Community Question Answering}.
  In \bibinfo{booktitle}{\emph{Proceedings of the 11th International Workshop
  on Semantic Evaluation, SemEval@ACL 2017, Vancouver, Canada, August 3-4,
  2017}}, \bibfield{editor}{\bibinfo{person}{Steven Bethard},
  \bibinfo{person}{Marine Carpuat}, \bibinfo{person}{Marianna Apidianaki},
  \bibinfo{person}{Saif~M. Mohammad}, \bibinfo{person}{Daniel~M. Cer}, {and}
  \bibinfo{person}{David Jurgens}} (Eds.). \bibinfo{publisher}{Association for
  Computational Linguistics}, \bibinfo{pages}{27--48}.
\newblock
\urldef\tempurl%
\url{https://doi.org/10.18653/v1/S17-2003}
\showDOI{\tempurl}


\bibitem[Peinelt et~al\mbox{.}(2020)]%
        {DBLP:conf/acl/PeineltNL20}
\bibfield{author}{\bibinfo{person}{Nicole Peinelt}, \bibinfo{person}{Dong
  Nguyen}, {and} \bibinfo{person}{Maria Liakata}.}
  \bibinfo{year}{2020}\natexlab{}.
\newblock \showarticletitle{tBERT: Topic Models and {BERT} Joining Forces for
  Semantic Similarity Detection}. In \bibinfo{booktitle}{\emph{Proceedings of
  the 58th Annual Meeting of the Association for Computational Linguistics,
  {ACL} 2020, Online, July 5-10, 2020}}, \bibfield{editor}{\bibinfo{person}{Dan
  Jurafsky}, \bibinfo{person}{Joyce Chai}, \bibinfo{person}{Natalie Schluter},
  {and} \bibinfo{person}{Joel~R. Tetreault}} (Eds.).
  \bibinfo{publisher}{Association for Computational Linguistics},
  \bibinfo{pages}{7047--7055}.
\newblock
\urldef\tempurl%
\url{https://doi.org/10.18653/v1/2020.acl-main.630}
\showDOI{\tempurl}


\bibitem[Pontes et~al\mbox{.}(2018)]%
        {DBLP:conf/taln/PontesHLT18}
\bibfield{author}{\bibinfo{person}{Elvys~Linhares Pontes},
  \bibinfo{person}{St{\'{e}}phane Huet}, \bibinfo{person}{Andr{\'{e}}a~Carneiro
  Linhares}, {and} \bibinfo{person}{Juan{-}Manuel Torres{-}Moreno}.}
  \bibinfo{year}{2018}\natexlab{}.
\newblock \showarticletitle{Predicting the Semantic Textual Similarity with
  Siamese {CNN} and {LSTM}}. In \bibinfo{booktitle}{\emph{Actes de la
  Conf{\'{e}}rence {TALN.} {CORIA-TALN-RJC} 2018 - Volume 1 - Articles longs,
  articles courts de TALN, Rennes, France, May 14-18, 2018}},
  \bibfield{editor}{\bibinfo{person}{Pascale S{\'{e}}billot} {and}
  \bibinfo{person}{Vincent Claveau}} (Eds.). \bibinfo{publisher}{{ATALA}},
  \bibinfo{pages}{311--320}.
\newblock
\urldef\tempurl%
\url{https://aclanthology.org/2018.jeptalnrecital-court.13/}
\showURL{%
\tempurl}


\bibitem[Rawat et~al\mbox{.}(2020)]%
        {DBLP:conf/bionlp/RawatWMRS20}
\bibfield{author}{\bibinfo{person}{Bhanu Pratap~Singh Rawat},
  \bibinfo{person}{Wei{-}Hung Weng}, \bibinfo{person}{So~Yeon Min},
  \bibinfo{person}{Preethi Raghavan}, {and} \bibinfo{person}{Peter Szolovits}.}
  \bibinfo{year}{2020}\natexlab{}.
\newblock \showarticletitle{Entity-Enriched Neural Models for Clinical Question
  Answering}. In \bibinfo{booktitle}{\emph{Proceedings of the 19th SIGBioMed
  Workshop on Biomedical Language Processing, BioNLP 2020, Online, July 9,
  2020}}, \bibfield{editor}{\bibinfo{person}{Dina Demner{-}Fushman},
  \bibinfo{person}{Kevin~Bretonnel Cohen}, \bibinfo{person}{Sophia Ananiadou},
  {and} \bibinfo{person}{Junichi Tsujii}} (Eds.).
  \bibinfo{publisher}{Association for Computational Linguistics},
  \bibinfo{pages}{112--122}.
\newblock
\urldef\tempurl%
\url{https://doi.org/10.18653/v1/2020.bionlp-1.12}
\showDOI{\tempurl}


\bibitem[Reimers and Gurevych(2019)]%
        {DBLP:conf/emnlp/ReimersG19}
\bibfield{author}{\bibinfo{person}{Nils Reimers} {and} \bibinfo{person}{Iryna
  Gurevych}.} \bibinfo{year}{2019}\natexlab{}.
\newblock \showarticletitle{Sentence-BERT: Sentence Embeddings using Siamese
  BERT-Networks}. In \bibinfo{booktitle}{\emph{Proceedings of the 2019
  Conference on Empirical Methods in Natural Language Processing and the 9th
  International Joint Conference on Natural Language Processing, {EMNLP-IJCNLP}
  2019, Hong Kong, China, November 3-7, 2019}},
  \bibfield{editor}{\bibinfo{person}{Kentaro Inui}, \bibinfo{person}{Jing
  Jiang}, \bibinfo{person}{Vincent Ng}, {and} \bibinfo{person}{Xiaojun Wan}}
  (Eds.). \bibinfo{publisher}{Association for Computational Linguistics},
  \bibinfo{pages}{3980--3990}.
\newblock
\urldef\tempurl%
\url{https://doi.org/10.18653/v1/D19-1410}
\showDOI{\tempurl}


\bibitem[Sakata et~al\mbox{.}(2019)]%
        {DBLP:conf/sigir/SakataSTK19}
\bibfield{author}{\bibinfo{person}{Wataru Sakata}, \bibinfo{person}{Tomohide
  Shibata}, \bibinfo{person}{Ribeka Tanaka}, {and} \bibinfo{person}{Sadao
  Kurohashi}.} \bibinfo{year}{2019}\natexlab{}.
\newblock \showarticletitle{{FAQ} Retrieval using Query-Question Similarity and
  BERT-Based Query-Answer Relevance}. In \bibinfo{booktitle}{\emph{{SIGIR}}}.
  \bibinfo{publisher}{{ACM}}, \bibinfo{pages}{1113--1116}.
\newblock


\bibitem[Severyn and Moschitti(2015)]%
        {DBLP:conf/sigir/SeverynM15}
\bibfield{author}{\bibinfo{person}{Aliaksei Severyn} {and}
  \bibinfo{person}{Alessandro Moschitti}.} \bibinfo{year}{2015}\natexlab{}.
\newblock \showarticletitle{Learning to Rank Short Text Pairs with
  Convolutional Deep Neural Networks}. In \bibinfo{booktitle}{\emph{Proceedings
  of the 38th International {ACM} {SIGIR} Conference on Research and
  Development in Information Retrieval, Santiago, Chile, August 9-13, 2015}},
  \bibfield{editor}{\bibinfo{person}{Ricardo Baeza{-}Yates},
  \bibinfo{person}{Mounia Lalmas}, \bibinfo{person}{Alistair Moffat}, {and}
  \bibinfo{person}{Berthier~A. Ribeiro{-}Neto}} (Eds.).
  \bibinfo{publisher}{{ACM}}, \bibinfo{pages}{373--382}.
\newblock
\urldef\tempurl%
\url{https://doi.org/10.1145/2766462.2767738}
\showDOI{\tempurl}


\bibitem[Sharma and et~al.(2011)]%
        {DBLP:conf/softcomp/SharmaC11}
\bibfield{author}{\bibinfo{person}{Sudhir~Kumar Sharma} {and}
  \bibinfo{person}{et al.}} \bibinfo{year}{2011}\natexlab{}.
\newblock \showarticletitle{An Adaptive Sigmoidal Activation Function Cascading
  Neural Networks}. In \bibinfo{booktitle}{\emph{SOCO}},
  Vol.~\bibinfo{volume}{87}. \bibinfo{pages}{105--116}.
\newblock


\bibitem[Sun et~al\mbox{.}(2020)]%
        {DBLP:conf/aaai/SunWLFTWW20}
\bibfield{author}{\bibinfo{person}{Yu Sun}, \bibinfo{person}{Shuohuan Wang},
  \bibinfo{person}{Yu{-}Kun Li}, \bibinfo{person}{Shikun Feng},
  \bibinfo{person}{Hao Tian}, \bibinfo{person}{Hua Wu}, {and}
  \bibinfo{person}{Haifeng Wang}.} \bibinfo{year}{2020}\natexlab{}.
\newblock \showarticletitle{{ERNIE} 2.0: {A} Continual Pre-Training Framework
  for Language Understanding}. In \bibinfo{booktitle}{\emph{The Thirty-Fourth
  {AAAI} Conference on Artificial Intelligence, {AAAI} 2020, The Thirty-Second
  Innovative Applications of Artificial Intelligence Conference, {IAAI} 2020,
  The Tenth {AAAI} Symposium on Educational Advances in Artificial
  Intelligence, {EAAI} 2020, New York, NY, USA, February 7-12, 2020}}.
  \bibinfo{publisher}{{AAAI} Press}, \bibinfo{pages}{8968--8975}.
\newblock
\urldef\tempurl%
\url{https://ojs.aaai.org/index.php/AAAI/article/view/6428}
\showURL{%
\tempurl}


\bibitem[Suzuki et~al\mbox{.}(2018)]%
        {DBLP:conf/paclic/SuzukiKSS18}
\bibfield{author}{\bibinfo{person}{Masaya Suzuki}, \bibinfo{person}{Kanako
  Komiya}, \bibinfo{person}{Minoru Sasaki}, {and} \bibinfo{person}{Hiroyuki
  Shinnou}.} \bibinfo{year}{2018}\natexlab{}.
\newblock \showarticletitle{Fine-tuning for Named Entity Recognition Using
  Part-of-Speech Tagging}. In \bibinfo{booktitle}{\emph{Proceedings of the 32nd
  Pacific Asia Conference on Language, Information and Computation, {PACLIC}
  2018, Hong Kong, December 1-3, 2018}},
  \bibfield{editor}{\bibinfo{person}{Stephen Politzer{-}Ahles},
  \bibinfo{person}{Yu{-}Yin Hsu}, \bibinfo{person}{Chu{-}Ren Huang}, {and}
  \bibinfo{person}{Yao Yao}} (Eds.). \bibinfo{publisher}{Association for
  Computational Linguistics}.
\newblock
\urldef\tempurl%
\url{https://aclanthology.org/Y18-1072/}
\showURL{%
\tempurl}


\bibitem[Vaswani et~al\mbox{.}(2017)]%
        {DBLP:conf/nips/VaswaniSPUJGKP17}
\bibfield{author}{\bibinfo{person}{Ashish Vaswani}, \bibinfo{person}{Noam
  Shazeer}, \bibinfo{person}{Niki Parmar}, \bibinfo{person}{Jakob Uszkoreit},
  \bibinfo{person}{Llion Jones}, \bibinfo{person}{Aidan~N. Gomez},
  \bibinfo{person}{Lukasz Kaiser}, {and} \bibinfo{person}{Illia Polosukhin}.}
  \bibinfo{year}{2017}\natexlab{}.
\newblock \showarticletitle{Attention is All you Need}. In
  \bibinfo{booktitle}{\emph{Advances in Neural Information Processing Systems
  30: Annual Conference on Neural Information Processing Systems 2017, December
  4-9, 2017, Long Beach, CA, {USA}}},
  \bibfield{editor}{\bibinfo{person}{Isabelle Guyon}, \bibinfo{person}{Ulrike
  von Luxburg}, \bibinfo{person}{Samy Bengio}, \bibinfo{person}{Hanna~M.
  Wallach}, \bibinfo{person}{Rob Fergus}, \bibinfo{person}{S.~V.~N.
  Vishwanathan}, {and} \bibinfo{person}{Roman Garnett}} (Eds.).
  \bibinfo{pages}{5998--6008}.
\newblock
\urldef\tempurl%
\url{https://proceedings.neurips.cc/paper/2017/hash/3f5ee243547dee91fbd053c1c4a845aa-Abstract.html}
\showURL{%
\tempurl}


\bibitem[Wan et~al\mbox{.}(2016)]%
        {DBLP:conf/aaai/WanLGXPC16}
\bibfield{author}{\bibinfo{person}{Shengxian Wan}, \bibinfo{person}{Yanyan
  Lan}, \bibinfo{person}{Jiafeng Guo}, \bibinfo{person}{Jun Xu},
  \bibinfo{person}{Liang Pang}, {and} \bibinfo{person}{Xueqi Cheng}.}
  \bibinfo{year}{2016}\natexlab{}.
\newblock \showarticletitle{A Deep Architecture for Semantic Matching with
  Multiple Positional Sentence Representations}. In
  \bibinfo{booktitle}{\emph{{AAAI}}}. \bibinfo{publisher}{{AAAI} Press},
  \bibinfo{pages}{2835--2841}.
\newblock


\bibitem[Wang et~al\mbox{.}(2017)]%
        {DBLP:conf/ijcai/WangHF17}
\bibfield{author}{\bibinfo{person}{Zhiguo Wang}, \bibinfo{person}{Wael Hamza},
  {and} \bibinfo{person}{Radu Florian}.} \bibinfo{year}{2017}\natexlab{}.
\newblock \showarticletitle{Bilateral Multi-Perspective Matching for Natural
  Language Sentences}. In \bibinfo{booktitle}{\emph{{IJCAI}}}.
  \bibinfo{publisher}{ijcai.org}, \bibinfo{pages}{4144--4150}.
\newblock


\bibitem[Xue et~al\mbox{.}(2008)]%
        {DBLP:conf/sigir/XueJC08}
\bibfield{author}{\bibinfo{person}{Xiaobing Xue}, \bibinfo{person}{Jiwoon
  Jeon}, {and} \bibinfo{person}{W.~Bruce Croft}.}
  \bibinfo{year}{2008}\natexlab{}.
\newblock \showarticletitle{Retrieval models for question and answer archives}.
  In \bibinfo{booktitle}{\emph{Proceedings of the 31st Annual International
  {ACM} {SIGIR} Conference on Research and Development in Information
  Retrieval, {SIGIR} 2008, Singapore, July 20-24, 2008}},
  \bibfield{editor}{\bibinfo{person}{Sung{-}Hyon Myaeng},
  \bibinfo{person}{Douglas~W. Oard}, \bibinfo{person}{Fabrizio Sebastiani},
  \bibinfo{person}{Tat{-}Seng Chua}, {and} \bibinfo{person}{Mun{-}Kew Leong}}
  (Eds.). \bibinfo{publisher}{{ACM}}, \bibinfo{pages}{475--482}.
\newblock
\urldef\tempurl%
\url{https://doi.org/10.1145/1390334.1390416}
\showDOI{\tempurl}


\bibitem[Yang et~al\mbox{.}(2019)]%
        {DBLP:conf/acl/YangZGJC19}
\bibfield{author}{\bibinfo{person}{Runqi Yang}, \bibinfo{person}{Jianhai
  Zhang}, \bibinfo{person}{Xing Gao}, \bibinfo{person}{Feng Ji}, {and}
  \bibinfo{person}{Haiqing Chen}.} \bibinfo{year}{2019}\natexlab{}.
\newblock \showarticletitle{Simple and Effective Text Matching with Richer
  Alignment Features}. In \bibinfo{booktitle}{\emph{Proceedings of the 57th
  Conference of the Association for Computational Linguistics, {ACL} 2019,
  Florence, Italy, July 28- August 2, 2019, Volume 1: Long Papers}},
  \bibfield{editor}{\bibinfo{person}{Anna Korhonen}, \bibinfo{person}{David~R.
  Traum}, {and} \bibinfo{person}{Llu{\'{\i}}s M{\`{a}}rquez}} (Eds.).
  \bibinfo{publisher}{Association for Computational Linguistics},
  \bibinfo{pages}{4699--4709}.
\newblock
\urldef\tempurl%
\url{https://doi.org/10.18653/v1/p19-1465}
\showDOI{\tempurl}


\bibitem[Zhang et~al\mbox{.}(2014)]%
        {DBLP:conf/cikm/ZhangWWLZ14}
\bibfield{author}{\bibinfo{person}{Kai Zhang}, \bibinfo{person}{Wei Wu},
  \bibinfo{person}{Haocheng Wu}, \bibinfo{person}{Zhoujun Li}, {and}
  \bibinfo{person}{Ming Zhou}.} \bibinfo{year}{2014}\natexlab{}.
\newblock \showarticletitle{Question Retrieval with High Quality Answers in
  Community Question Answering}. In \bibinfo{booktitle}{\emph{Proceedings of
  the 23rd {ACM} International Conference on Conference on Information and
  Knowledge Management, {CIKM} 2014, Shanghai, China, November 3-7, 2014}},
  \bibfield{editor}{\bibinfo{person}{Jianzhong Li},
  \bibinfo{person}{Xiaoyang~Sean Wang}, \bibinfo{person}{Minos~N. Garofalakis},
  \bibinfo{person}{Ian Soboroff}, \bibinfo{person}{Torsten Suel}, {and}
  \bibinfo{person}{Min Wang}} (Eds.). \bibinfo{publisher}{{ACM}},
  \bibinfo{pages}{371--380}.
\newblock
\urldef\tempurl%
\url{https://doi.org/10.1145/2661829.2661908}
\showDOI{\tempurl}


\bibitem[Zhou et~al\mbox{.}(2011)]%
        {DBLP:conf/acl/ZhouCZL11}
\bibfield{author}{\bibinfo{person}{Guangyou Zhou}, \bibinfo{person}{Li Cai},
  \bibinfo{person}{Jun Zhao}, {and} \bibinfo{person}{Kang Liu}.}
  \bibinfo{year}{2011}\natexlab{}.
\newblock \showarticletitle{Phrase-Based Translation Model for Question
  Retrieval in Community Question Answer Archives}. In
  \bibinfo{booktitle}{\emph{The 49th Annual Meeting of the Association for
  Computational Linguistics: Human Language Technologies, Proceedings of the
  Conference, 19-24 June, 2011, Portland, Oregon, {USA}}},
  \bibfield{editor}{\bibinfo{person}{Dekang Lin}, \bibinfo{person}{Yuji
  Matsumoto}, {and} \bibinfo{person}{Rada Mihalcea}} (Eds.).
  \bibinfo{publisher}{The Association for Computer Linguistics},
  \bibinfo{pages}{653--662}.
\newblock
\urldef\tempurl%
\url{https://aclanthology.org/P11-1066/}
\showURL{%
\tempurl}


\end{thebibliography}

\appendix
\end{CJK}
\end{document}